\documentclass{svproc}
\usepackage{url}

\usepackage{amsmath,amsfonts,amssymb}
\usepackage{stmaryrd}
\usepackage{graphicx}
\usepackage[hidelinks]{hyperref}
\usepackage[utf8]{inputenc}
\usepackage[small]{caption}
\usepackage{cleveref}
\usepackage[font=footnotesize]{subcaption}
\usepackage[normalem]{ulem}
\usepackage{multicol}
\usepackage{multirow}
\usepackage{soul}
\usepackage{color}

\begin{document}
\mainmatter              % start of a contribution
%
% \title{Small-world networks for collective learning\\ in multi-agent systems}
\title{The Impact of Network Connectivity\\ on Collective Learning}
% \title{Connectivity assumptions in multi-agent systems}
% Well-mixed vs. well-connected
%
\titlerunning{The Impact of Network Connectivity on Collective Learning}  % abbreviated title (for running head)
%                                     also used for the TOC unless
%                                     \toctitle is used
%
\author{Michael Crosscombe \and Jonathan Lawry}
% Michael Crosscombe\inst{1} \and Jonathan Lawry\inst{1}
% \and David Harvey\inst{2}}
%
\authorrunning{M. Crosscombe and J. Lawry} % abbreviated author list (for running head)
%
%%%% list of authors for the TOC (use if author list has to be modified)
% \tocauthor{Ivar Ekeland, Roger Temam, Jeffrey Dean, David Grove, Craig Chambers, Kim B. Bruce, and Elisa Bertino}
%
\institute{Department of Engineering Mathematics, University of Bristol\\ Bristol BS8 1UB, United Kingdom\\
\email{\{m.crosscombe,j.lawry\}@bristol.ac.uk}}%,\\ WWW home page:
% \texttt{http://users/\homedir iekeland/web/welcome.html}
% \and
% Research Technology and Innovation, Thales Group, United Kingdom\\
% \email{david.harvey@uk.thalesgroup.com}
% }

% \institute{Princeton University, Princeton NJ 08544, USA,\\
% \email{I.Ekeland@princeton.edu},\\ WWW home page:
% \texttt{http://users/\homedir iekeland/web/welcome.html}
% \and
% Universit\'{e} de Paris-Sud,
% Laboratoire d'Analyse Num\'{e}rique, B\^{a}timent 425,\\
% F-91405 Orsay Cedex, France}

\maketitle              % typeset the title of the contribution

\begin{abstract}
In decentralised autonomous systems it is the interactions between individual agents which govern the collective behaviours of the system. These local-level interactions are themselves often governed by an underlying network structure.
These networks are particularly important for collective learning and decision-making whereby agents must gather evidence from their environment and propagate this information to other agents in the system.
Models for collective behaviours may often rely upon the assumption of total connectivity between agents to provide effective information sharing within the system, but this assumption may be ill-advised.
In this paper we investigate the impact that the underlying network has on performance in the context of collective learning. Through simulations we study small-world networks with varying levels of connectivity and randomness and conclude that totally-connected networks result in higher average error when compared to networks with less connectivity. Furthermore, we show that networks of high regularity outperform networks with increasing levels of random connectivity.
% We would like to encourage you to list your keywords within
% the abstract section using the \keywords{...} command.
\end{abstract}
\section{Introduction}
\label{sec:introduction}

Reasoning about collective behaviours in autonomous systems can be difficult when the system-level behaviour emerges from local-level interactions, i.e., between individual agents.
Due to the complexity of such systems we often rely on a series of modelling assumptions to effectively reason about their resulting dynamics.
One such assumption is the ``well-stirred system'' assumption, which we restate to be the following: In a well-stirred system each agent is equally likely to encounter any other agent in the system and, therefore, each interaction is regarded as an independent event~\cite{Parker}. In other words, in a well-stirred system we treat agents as nodes in a totally-connected network or complete graph and the stochastic interactions of agents can hence be modelled by selecting edges at random from the network.
While this assumption captures the notion that agents conducting random walks are likely to encounter one another at random, e.g., whilst exploring an environment, the impact of this assumption is not well-understood. Alternative network structures, such as small-world networks~\cite{Watts1998}, have been studied in both artificial and biological systems~\cite{Newman,Masuda} and provide a means of varying the connectivity of the network while preserving desirable properties such as short, consistent path-lengths.

% In this paper we investigate the impact of the underlying network structure on a collective learning problem in multi-agent systems.
In this paper we will study consensus formation in a collective learning setting, in which agents learn both from evidence gathered directly from the environment as well as indirectly from one another, through a process of belief fusion. In this context we shall investigate the impact of the underlying network structure on collective learning by studying small-world networks with varying degrees of connectivity and stochasticity.
The rest of this paper is structured as follows: In \Cref{sec:related-work} we provide an overview of the relevant literature on various formulations of collective learning as well as small-world networks. In \Cref{sec:model} we describe a propositional approach to collective learning in which agents attempt to learn a full state description of the environment, followed by a brief overview of small-world networks in \Cref{sec:network}. Then, in \Cref{sec:experiments} we detail simulation experiments in which we study small-world networks of varying degrees of connectivity and random rewiring. Finally, in \Cref{sec:conclusion} we close with some discussions and conclusions.

\section{Related work}
\label{sec:related-work}

Consensus formation within a large group of individuals has long been studied in the form of `opinion pooling' since the works of Stone~\cite{Stone} and DeGroot~\cite{DeGroot}. These early works considered the (usually linear) weighted fusion of beliefs between `experts' initialised with prior beliefs and without considering the influence of external evidence.
Instead of learning a description of the world, models of opinion dynamics are typically concerned with a single proposition of contention and beliefs are then denoted by a single real value that represents an agent's certainty in the truth or falsity of the proposition~\cite{Hegselmann2002}.
Agents simply conduct weighted combinations of their beliefs under interaction limitations imposed by concepts such as bounded confidence~\cite{Hegselmann2005,Dabarera2019}.
Three-valued representations of beliefs have also been studied in the context of opinion dynamics in \cite{Balenzuela}, which thresholds the underlying real value into three states, and in the context of consensus formation in \cite{Perron}, which proposes a three-valued fusion operator that assigns the third truth state to resolve conflict between two strictly opposing opinions.

Collective learning is the combination of two distinct processes: belief fusion between agents; and the updating of beliefs based on evidence gathered from the environment.
% The latter process is what distinguishes collective learning from other forms of consensus formation.
The interaction of these two processes in the context of social epistemology has been explored in~\cite{Douven2011}, in which the argument is made that communication between agents not only acts to propagate information more effectively through the system but also provides an error-correcting effect when the evidence being gathered may be erroneous. Further studies of this effect in multi-agent systems and swarm robotics can be found in \cite{Crosscombe2016,Lee2018,Douven2019}. In a robotics setting this evidence may take the form of signals received by the robots' on-board sensors. For example, in \cite{Valentini2015} experiments are conducted in which Kilobots use their ambient light sensors to determine the quality (i.e., light intensity) of a particular location~\cite{Rubenstein}.

Due to recent developments in multi-robot systems and swarm robotics, the increasing viability of their deployment outside of the lab has led to a surge of interest in the development and understanding of collective behaviours~\cite{Brambilla,Schranz}. Many of the current solutions are based in nature, e.g., the house-hunting behaviours studied in ants and honeybees~\cite{Franks} are instances of the best-of-$n$ problem~\cite{Parker2011,Valentini2017}. It is in this context that \cite{Parker} first proposed the adoption of the well-stirred system assumption which is to be found, implicitly or explicitly, in many recent models for collective learning~\cite{Crosscombe2017,Lee2018,Lawry2019}.
However, there has been increasing effort to understand the impact of the underlying network topology on collective behaviours in multi-agent systems. In \cite{Olfati-Saber} the authors compare the effects that different network topologies, including small-world networks~\cite{Watts1998} and random networks~\cite{Erdos1959}, have on the consensus formation process. In the context of consensus formation, Baronchelli~\cite{Baronchelli} investigates convergence dynamics for various network topologies.

\section{A propositional model for collective learning}
\label{sec:model}

Consider a collective learning problem whereby a population of agents are attempting to reach a consensus about the true state of their environment. Let us suppose that such an environment can be described by a set of $n$ propositional variables $\mathcal{P} = \{p_1,\dots,p_n\}$ and that an agent's belief about the environment (i.e., a possible world) is an assignment of truth values to each of the propositional variables denoted by $b : \mathcal{P} \rightarrow \{0,\frac{1}{2},1\}^n$. Here we adopt a three-valued propositional logic with the truth values $0$ and $1$ corresponding to false and true, respectively, while the third truth value $\frac{1}{2}$ denotes \emph{unknown}. This additional truth value allows for agents to express uncertainty in their beliefs about the world. For notational convenience, let us represent an agent's belief $B$ by the $n$-tuple $\langle B(p_1), \dots, B(p_n) \rangle$. Then an agent's belief may express uncertainty about the world by the assignment of the truth value $\frac{1}{2}$ to any of the propositional variables in $\mathcal{P}$. For example, for $n = 2$ propositional variables the belief $B = \langle 1, 0 \rangle$ expresses an agent's belief that $p_1$ is true while $p_2$ is false. We can say that this belief expresses absolute certainty in the state of the world. Alternatively, the belief $B^\prime = \langle \frac{1}{2}, \frac{1}{2} \rangle$ expresses an agent's uncertainty about both of the propositions $p_1$ and $p_2$, thus indicating the agent's total lack of certainty regarding the state of the world.

\begin{table}[t]
    \centering
    \caption{The fusion operator $\odot$ applied to beliefs $B$ and $B^\prime$.}
    \vspace{1em}
    \setlength{\tabcolsep}{6pt}
    \def\arraystretch{1.4}
    \begin{tabular}{ccccc}
        & \multicolumn{4}{l}{\hspace*{0.65cm}$B^\prime(p_i)$} \\ \cline{2-5} 
        \multicolumn{1}{l|}{\multirow{4}{*}{$B(p_i)$}} &
        % First row
        \multicolumn{1}{l|}{$\odot$} &
        \multicolumn{1}{l|}{0} &
        \multicolumn{1}{l|}{$\frac{1}{2}$} &
        \multicolumn{1}{l|}{1} \\ \cline{2-5} 
        % Second row
        \multicolumn{1}{l|}{} &
        \multicolumn{1}{l|}{0} &
        \multicolumn{1}{l|}{0} &
        \multicolumn{1}{l|}{0} &
        \multicolumn{1}{l|}{$\frac{1}{2}$} \\ \cline{2-5}
        % Third row
        \multicolumn{1}{l|}{} &
        \multicolumn{1}{l|}{$\frac{1}{2}$} &
        \multicolumn{1}{l|}{0}&
        \multicolumn{1}{l|}{$\frac{1}{2}$} &
        \multicolumn{1}{l|}{1} \\ \cline{2-5}
        % Fourth row
        \multicolumn{1}{l|}{}&
        \multicolumn{1}{l|}{1} &
        \multicolumn{1}{l|}{$\frac{1}{2}$} &
        \multicolumn{1}{l|}{1} &
        \multicolumn{1}{l|}{1} \\ \cline{2-5} 
    \end{tabular}
    \label{tab:fusion}
\end{table}

We propose to combine the beliefs of agents in a pairwise manner as follows: the fusion operator $\odot$ is a binary operator defined on $\{0,\frac{1}{2},1\}$ as given in \Cref{tab:fusion}. This operator is applied element-wise to all of the propositional variables $p_i$ in $\mathcal{P}$ so that, given two beliefs $B$ and $B^\prime$ corresponding to the beliefs of two agents, the fused belief is given by
\begin{gather}
    B \odot B^\prime = \langle B(p_1) \odot B^\prime(p_1), \dots, B(p_n) \odot B^\prime(p_n) \rangle.
\end{gather}
A pairwise consensus is thus formed by both agents adopting the fused belief $B \odot B^\prime$.
% Highlight the operator's intended consequence of resolving inconsistencies while forcing consensus

Evidential updating is the process by which an agent selects a proposition (e.g., a feature of its environment) to investigate and, upon receiving evidence, updates its belief to reflect this evidence. Firstly, to decide which proposition to investigate an agent selects a single proposition at random from the set of propositions about which they are uncertain, i.e., where $B(p_i) = \frac{1}{2}$.
Having selected a proposition $p_i$ to investigate, an agent then receives evidence with probability $r$ or learns nothing with probability $1 - r$, where $r$ is an evidence rate quantifying the sparsity of evidence in the environment.
Evidence takes the form of an assertion about the truth value of the chosen proposition $p_i$ as follows: $E = \langle \frac{1}{2}, \dots, S^*(p_i), \dots, \frac{1}{2} \rangle$ where $S^* : \mathcal{P} \rightarrow \{0,1\}^n$ denotes the true state of the world.
Secondly, upon gathering evidence $E$, the agent then updates its belief $B$ to $B|E$ using the same fusion operator given in \Cref{tab:fusion} such that
\begin{align}
    \begin{split}
        B|E &= \langle B(p_1) \odot E(p_1), \dots, B(p_i) \odot E(p_i), \dots, B(p_n) \odot E(p_n) \rangle\\
        &= B \odot E.
    \end{split}
\end{align}
Notice that we are also using the fusion operator $\odot$ to update beliefs based on evidence and that updating in this manner does not therefore alter the truth values for the propositions $p_j \in \mathcal{P}$ where $p_j \neq p_i$ because $E(p_j) = \frac{1}{2}$.
An agent repeats this process of gathering evidence until the set of propositions about which it is uncertain is empty, or rather that it holds a belief of total certainty, at which point it chooses to stop looking for evidence. Also notice that while evidential updating in this manner can only lead to agents becoming more certain in their beliefs, the process of agents combining their beliefs via the fusion operator $\odot$ can also lead to agents becoming more uncertain when the fusing agents disagree about the truth value of a given proposition. For example, supposing that $B_1(p_i) = 1$ and $B_2(p_i) = 0$, then upon the agents fusing their beliefs such that $B_1 \odot B_2(p_i) = \frac{1}{2}$, both agents will attempt to seek additional evidence about proposition $p_i$, either directly from the environment or indirectly via fusion with other agents, having become uncertain about the truth value of $p_i$.

Let us now assume that the evidence gathering process may be noisy (e.g., due to sensor noise or a noisy environment). Evidence shall then take the following form:
\begin{gather}\label{eq:noisy-evidence}
    E(p_i) = 
    \begin{cases}
        ~ S^*(p_i) &: ~\text{ with probability }1 - \epsilon \\
        ~ 1 -  S^*(p_i) &: ~\text{ with probability } \epsilon
    \end{cases}
\end{gather}
where $\epsilon \in [0,0.5]$ is a noise parameter denoting the probability that the evidence is erroneous.

To measure the performance of a given population in the context of a collective learning problem we introduce a measure of the average error of the population.

\begin{definition} Average error\\
\label{def:average-error}
    The average error of a population of $m$ agents is the normalised difference between each agent's belief $B$ and the true state of the world $S^*$ averaged across the population as follows:
    \begin{gather*}
        \frac{1}{m} \frac{1}{n} \sum_{j = 1}^m \sum_{i=1}^n \left| B_j(p_i) - S^*(p_i) \right|.
    \end{gather*}
\end{definition}

As mentioned previously, we often adopt the well-stirred system assumption which states that an interaction between any two agents is an independent event and is therefore equally likely for any pair of agents in the population~\cite{Parker}.
In the network agents are represented by nodes and the existence of edges between them represents the ability of the agents to communicate directly with one another.
In networks with less-than-total connectivity the lack of an edge between two agents means that they cannot communicate directly, although information may still be shared via other agents in the population through the process of belief fusion. In the following section we introduce small-world networks to study the impact that the underlying network structure has on the collective learning process and to challenge the well-stirred system assumption.

\section{Small-world networks}
\label{sec:network}

\begin{figure*}[t]
\begin{subfigure}{.23\textwidth}
\begin{center}
    \includegraphics[width=0.75\textwidth]{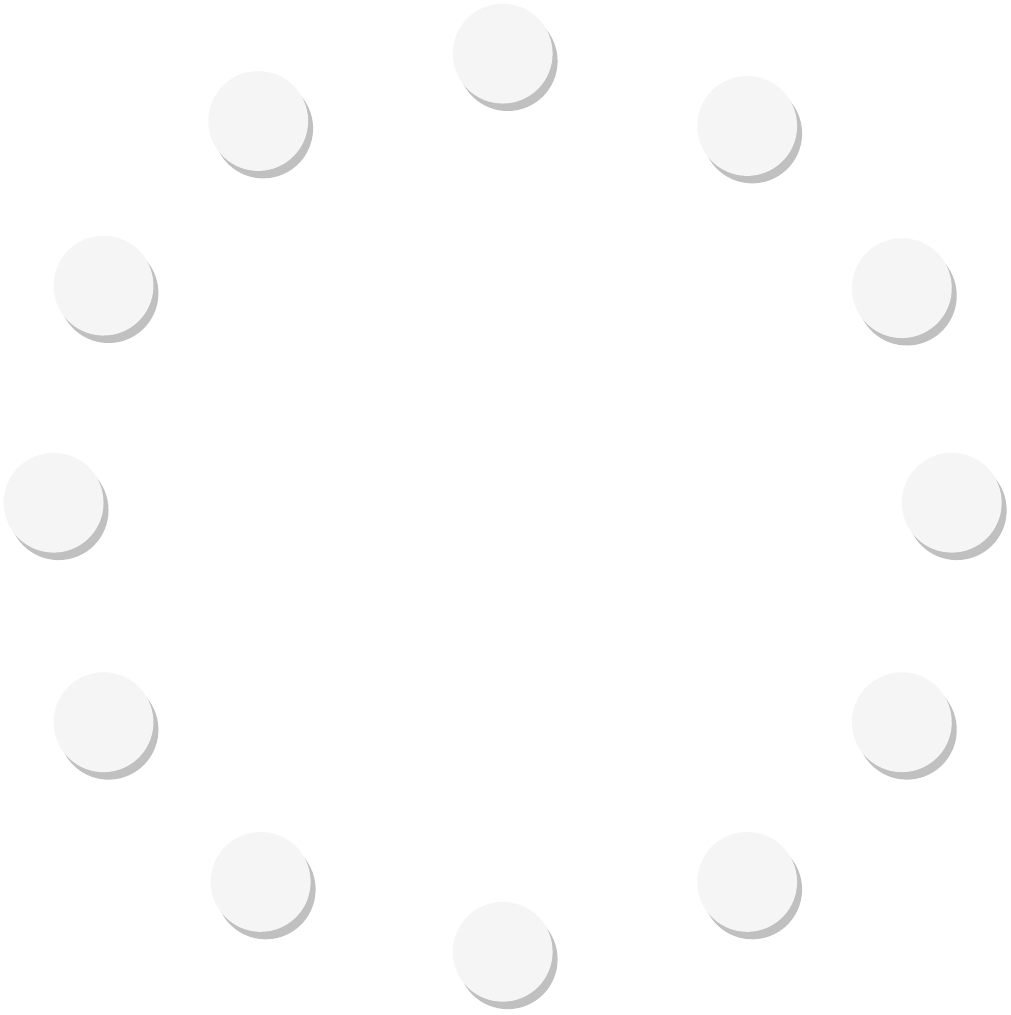}
    \subcaption{}
    \label{diag:small-world-1}
\end{center}
\end{subfigure}\hfill
\begin{subfigure}{.23\textwidth}
\begin{center}
    \includegraphics[width=0.75\textwidth]{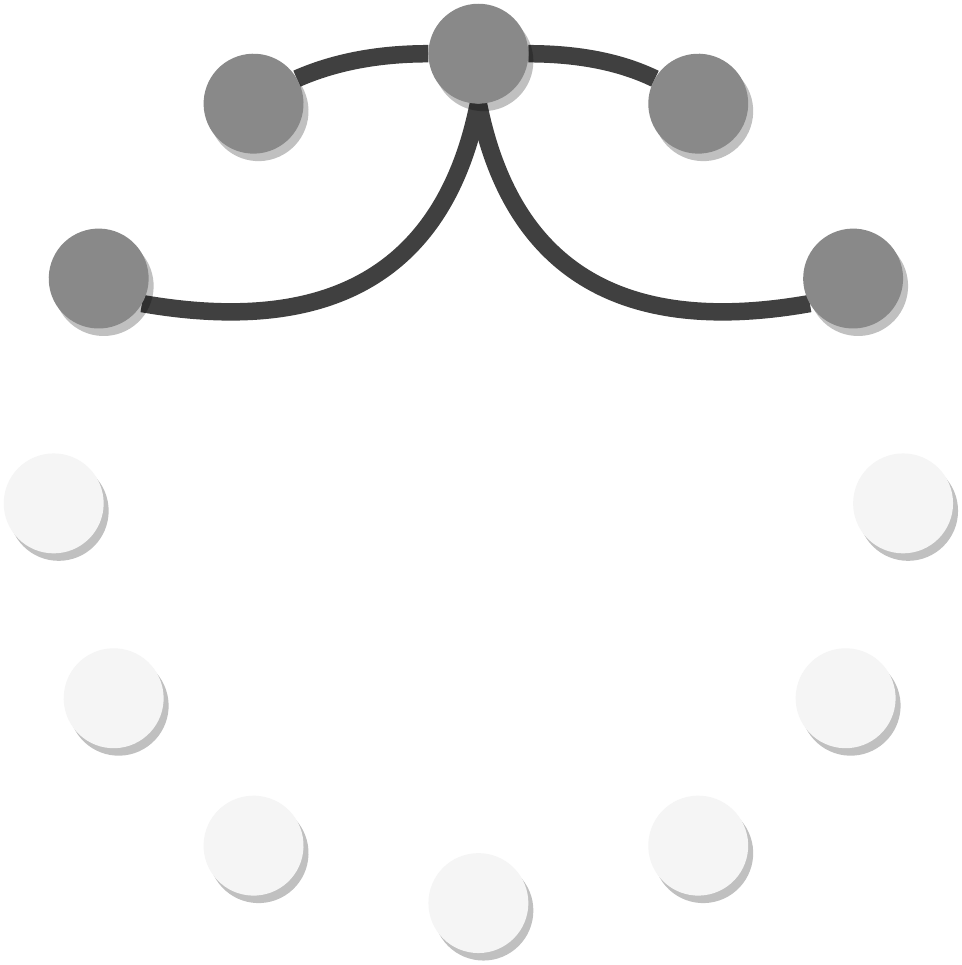}
    \subcaption{}
    \label{diag:small-world-2}
\end{center}
\end{subfigure}\hfill
\begin{subfigure}{.23\textwidth}
\begin{center}
    \includegraphics[width=0.75\textwidth]{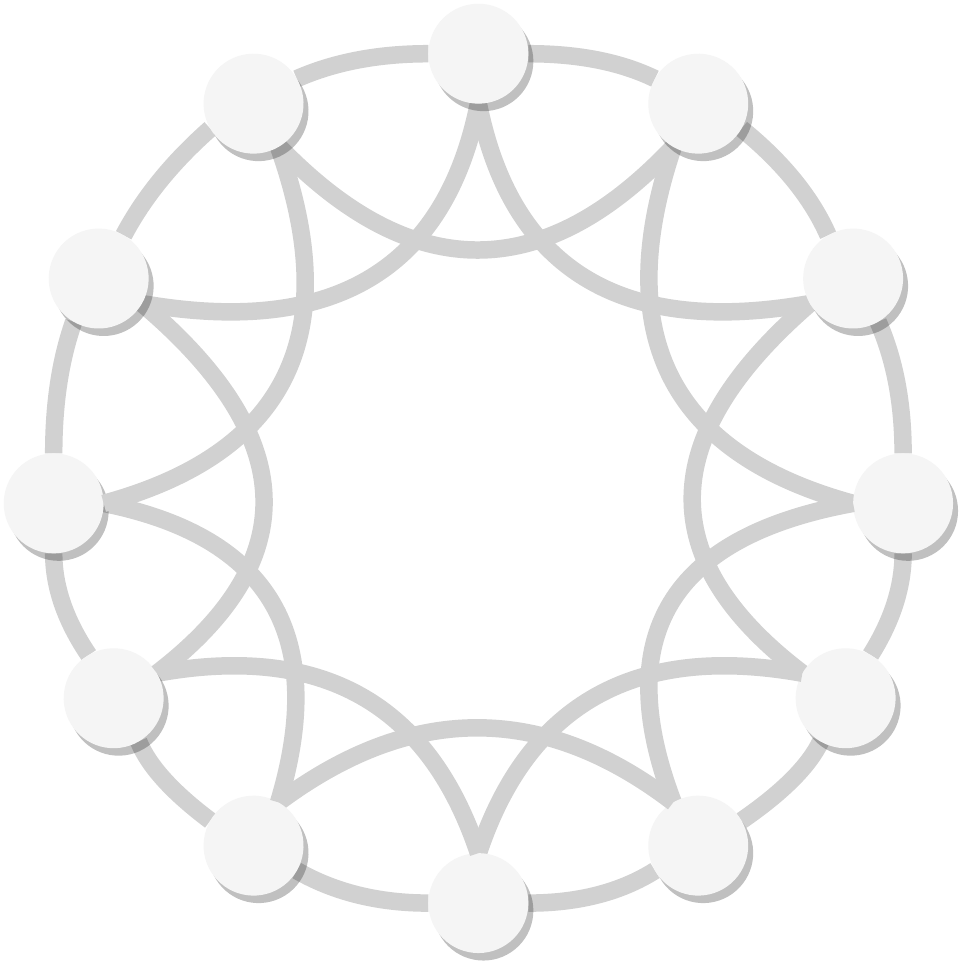}
    \subcaption{}
    \label{diag:small-world-3}
\end{center}
\end{subfigure}\hfill
\begin{subfigure}{.23\textwidth}
\begin{center}
    \includegraphics[width=0.75\textwidth]{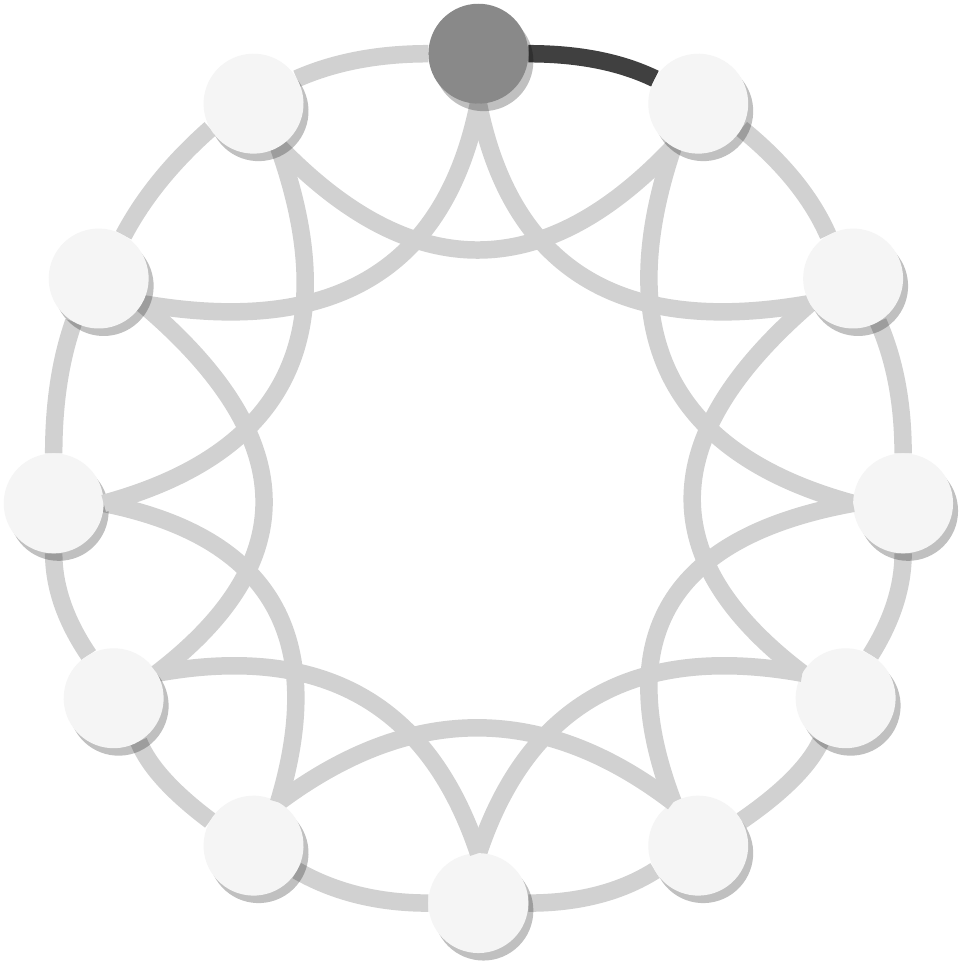}
    \subcaption{}
    \label{diag:small-world-4}
\end{center}
\end{subfigure}\hfill

\begin{subfigure}{.23\textwidth}
\begin{center}
    \includegraphics[width=0.75\textwidth]{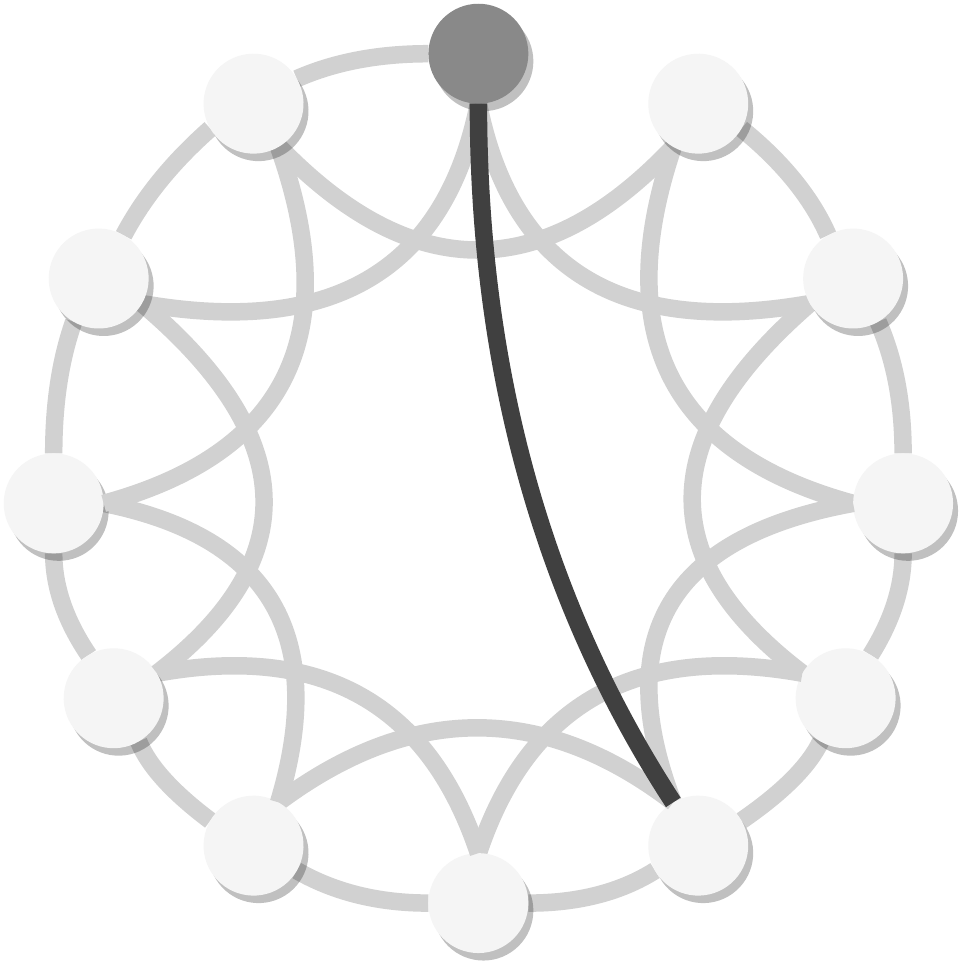}
    \subcaption{}
    \label{diag:small-world-5}
\end{center}
\end{subfigure}\hfill
\begin{subfigure}{.23\textwidth}
\begin{center}
    \includegraphics[width=0.75\textwidth]{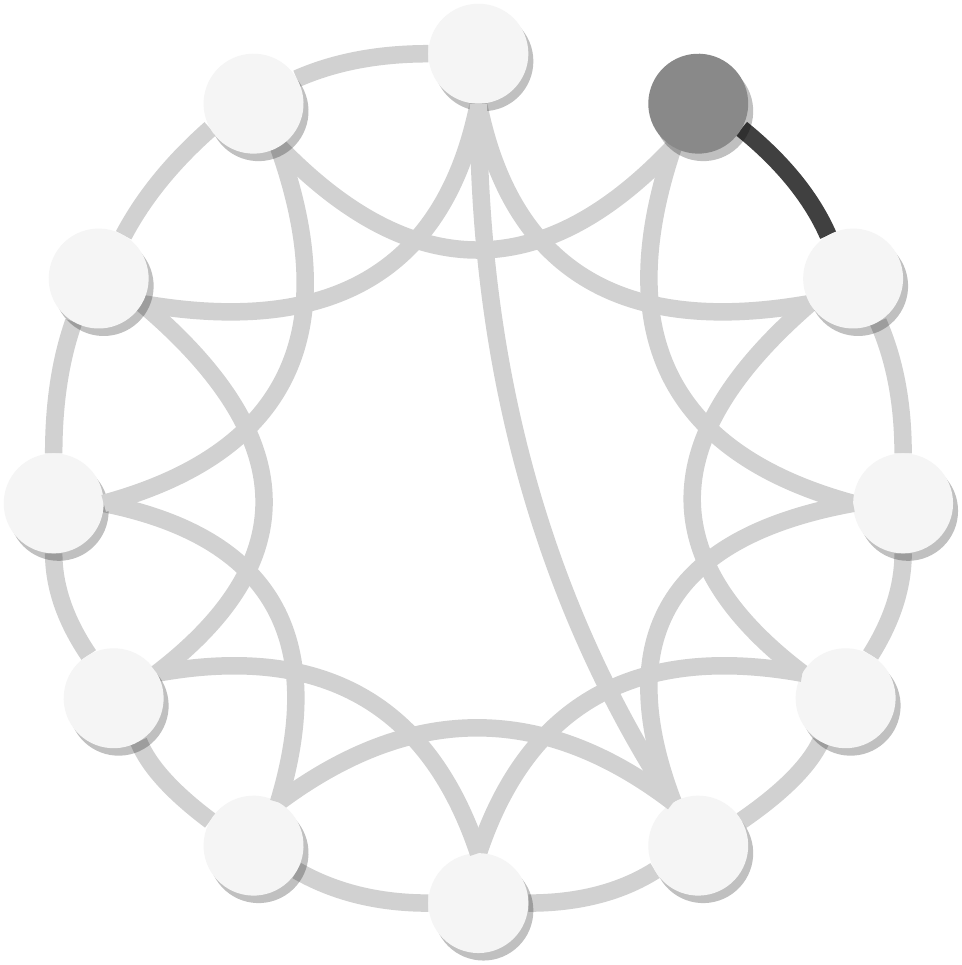}
    \subcaption{}
    \label{diag:small-world-6}
\end{center}
\end{subfigure}\hfill
\begin{subfigure}{.23\textwidth}
\begin{center}
    \includegraphics[width=0.75\textwidth]{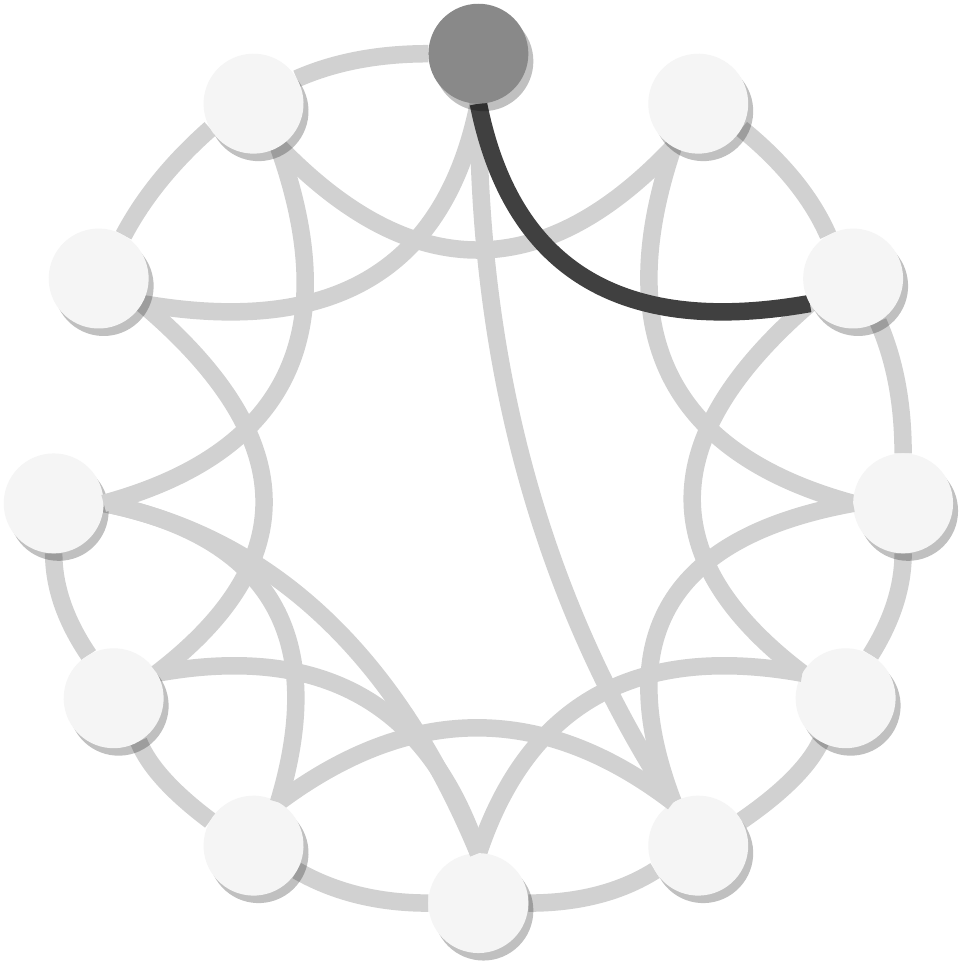}
    \subcaption{}
    \label{diag:small-world-7}
\end{center}
\end{subfigure}\hfill
\begin{subfigure}{.23\textwidth}
\begin{center}
    \includegraphics[width=0.75\textwidth]{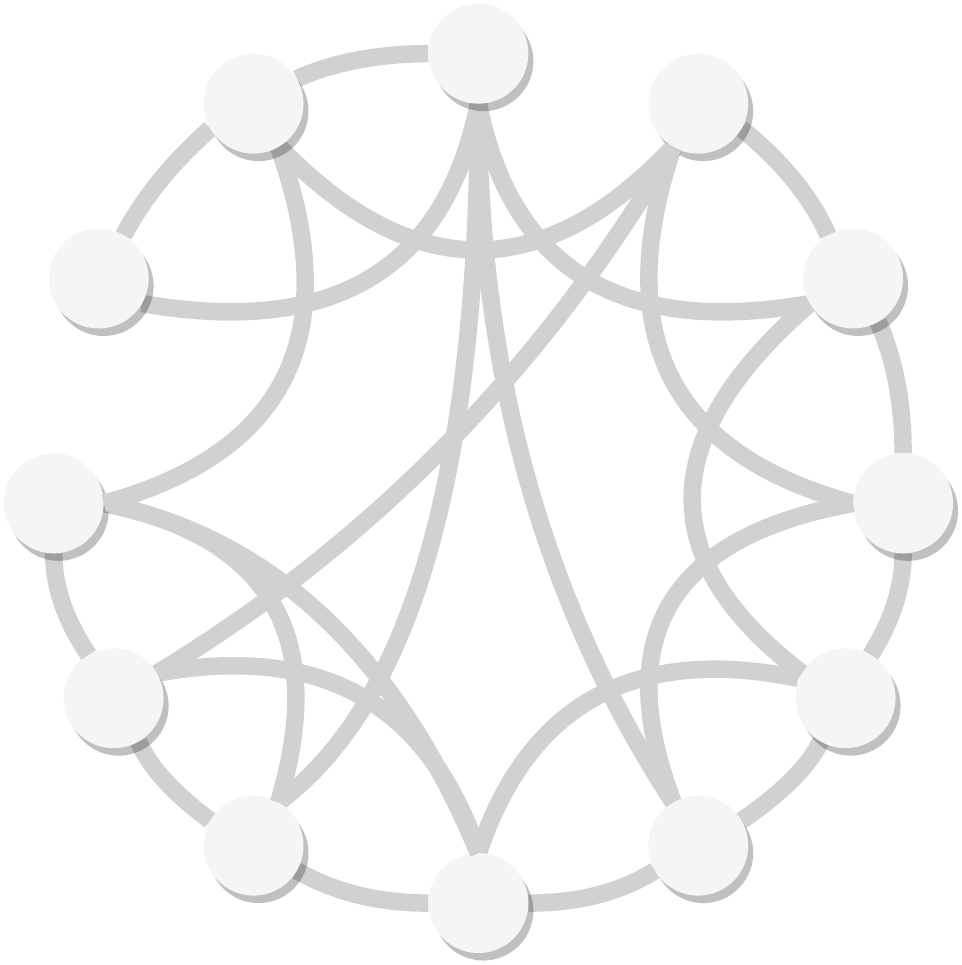}
    \subcaption{}
    \label{diag:small-world-8}
\end{center}
\end{subfigure}\hfill
\caption{The generation of a Watts-Strogatz small-world graph illustrated for $m = 12, k = 4$ and $\rho = 0.2$.}
\label{diag:small-world-graph}
\end{figure*}

The network topology of many real-world systems often lies somewhere between total regularity and total randomness, as discussed by Watts and Strogatz in their seminal work introducing small-world networks~\cite{Watts1998,Watts2003}. Indeed the concept of a small-world network -- in which nodes are connected to their $k$ nearest neighbouring nodes -- more accurately reflects the kinds of networks that emerge in both natural and artificial self-organising systems, with examples in social networks~\cite{Newman} and neural networks~\cite{Masuda}. Many social systems, both biological and engineered, exhibit these small-world dynamics due to their spatial properties. For example in swarm robotics, where each embodied agent is typically composed of low-cost hardware and consequently possesses a limited radius within which it can communicate~\cite{Rubenstein}, the distance between agents determines the connectivity of the underlying communication network which will often resemble a small-world network.

Small-world networks are parameterised by two variables $k$ and $\rho$, where $k$ denotes the number of nearest neighbours to which each node in the graph is connected and $\rho$ denotes the probability of rewiring an existing edge to a different agent. A small-world graph, as illustrated in~\Cref{diag:small-world-graph}, is generated in the following way: (a) begin with $n$ vertices ordered in a ring; (b) for each vertex in the graph, connect it to its $k$ nearest neighbours until; (c) a regular small-world graph is formed. (d) For each vertex (moving clockwise around the ring), select the edge connected to its clockwise-nearest neighbour; (e) with probability $\rho$ reconnect that edge to another vertex selected uniformly at random, unless doing so produces a duplicate edge. (f) Continue this process clockwise for all vertices. (g) Repeat this process for their second-nearest clockwise neighbour until; (h) each original edge in the graph has been considered once. Notice that a small-world network with $k = (m - 1)$ is equivalent to a totally-connected network in which each agent is connected to every other agent.

In the following section we describe simulation experiments applying the model in \Cref{sec:model} to small-world networks with varying degrees of connectivity and regularity.

\section{Agent-based simulations}
\label{sec:experiments}

We study a collective learning scenario in which the environment can be described by $n = 100$ propositions. Without loss of generality we define the true state of the world $S^*$ to be $\langle 1, 1, 1, \dots, 1 \rangle$. For each experiment we initialise a population of $k = 100$ agents holding totally ignorant beliefs, i.e., at time step $t = 0$ each agent holds the belief $B(p_i) = \frac{1}{2}$ for $i = 1, \dots, 100$. By \Cref{def:average-error} the average error of such a belief, representing complete uncertainty, is $0.5$ and therefore each population shall begin with an average error of $0.5$. Furthermore, should a population converge on the true state of the world $S^*$ then the average error shall be $0$.
A population-wide evidence rate $r \in (0,1]$ determines the frequency with which agents successfully obtain evidence from the environment. In other words, during each time step every agent has an equal probability $r$ of updating their belief based on evidence. As described previously, each piece of evidence pertains to a single proposition about which the investigating agent is uncertain, i.e., where $B(p_i) = \frac{1}{2}$.
This evidence is also likely to be noisy with $\epsilon \in [0,0.5]$ denoting the probability that a piece of evidence is incorrect. Notice that for $\epsilon = 0.5$ the evidence becomes random with an equal probability of learning that the investigated proposition is either true or false.
Finally, in addition to evidential updating, at each time step one edge in the graph is selected at random and the connected pair of agents combine their beliefs using the fusion operator defined in \Cref{tab:fusion}.

For a given set of parameter values we average the results over $100$ independent runs and each run takes a maximum of $10,000$ time steps, or until convergence occurs. Here we define convergence as the beliefs of the population remaining unchanged for $100$ interactions, where an interaction is updating either based on evidence or on the beliefs of other agents, i.e., via fusion. For line plots, the shaded regions represent $10^{\text{th}}$ and $90^{\text{th}}$ percentiles.

\subsection{Convergence results for regular small-world networks}
\label{ssec:results}

\begin{figure*}[t]
\begin{subfigure}{1\textwidth}
\centering
    \includegraphics[width=0.8\textwidth]{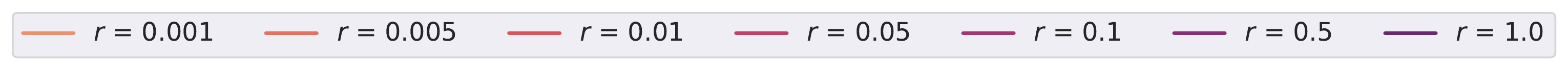}
\end{subfigure}\hfill
\centering
\begin{subfigure}{0.4\textwidth}
    \includegraphics[width=1\textwidth]{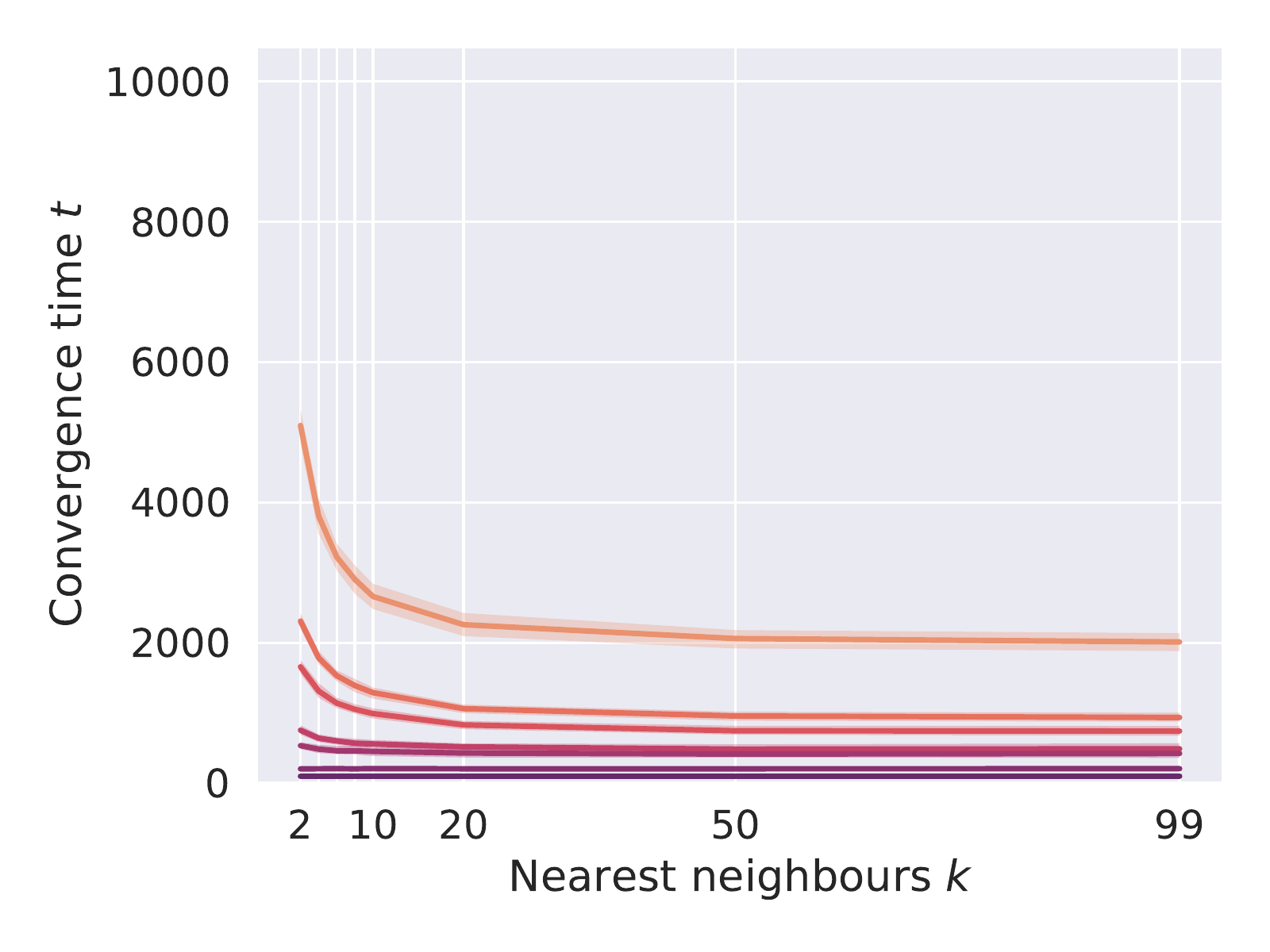}
    \subcaption{Noise $\epsilon = 0.0$.}
    \label{fig:time-noise-0}
\end{subfigure}\hspace{2em}
\begin{subfigure}{0.4\textwidth}
    \includegraphics[width=1\textwidth]{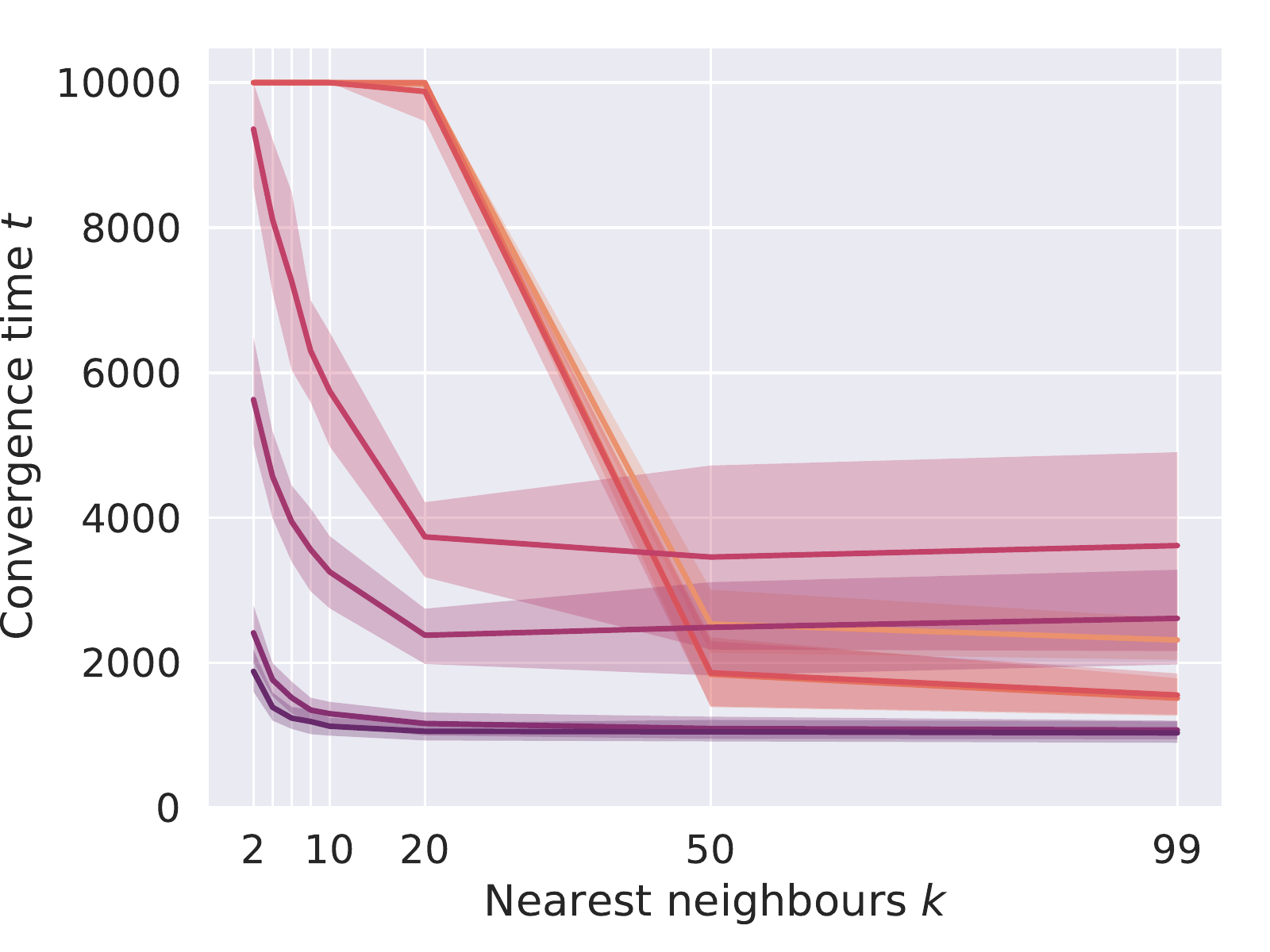}
    \subcaption{Noise $\epsilon = 0.3$.}
    \label{fig:time-noise-0.3}
\end{subfigure}\hfill
% \begin{subfigure}{.33\textwidth}
%     \includegraphics[width=1\textwidth]{{time_WS_100_states_100_agents_0.00_con_0.50_noise_shaded}.pdf}
%     \subcaption{Noise $\epsilon = 0.5$.}
%     \label{fig:time-noise-0.5}
% \end{subfigure}\hfill
\caption{Convergence time for a population of $100$ agents across nearest neighbours $k$. Each line depicts a different evidence rate $r \in [0.01,1]$.}
\label{fig:convergence-time}
\end{figure*}

In this section we show convergence results for regular small-world networks without random rewiring (i.e., where $\rho = 0$) as depicted in \Cref{diag:small-world-3}.
\Cref{fig:convergence-time} shows the average convergence time across different networks with nearest neighbours $k$ for different levels of noise $\epsilon$. Each solid line represents a different evidence rate $r$ between $0.001$ and $1$. For $\epsilon = 0$ in \Cref{fig:time-noise-0} we depict a noise-free scenario in which evidence is always accurate. Broadly, we see that time to convergence decreases as the network connectivity $k$ increases and that a totally-connected network results in the fastest convergence times for all evidence rates $r$. Additionally, we see that the average convergence time decreases as the evidence rate $r$ increases. This is to be expected as the greater the frequency with which the population receives evidence, the faster the agents learn about their environment.
In \Cref{fig:time-noise-0.3} we see that for a moderately noisy environment with $\epsilon = 0.3$ and lower evidence rates $r \leq 0.01$ the population no longer converges in under $10,000$ time steps for networks of connectivity $k < 50$. For networks with greater connectivity, i.e., $k \geq 50$, the population once again reaches a steady state in under $3,000$ time steps on average. For higher evidence rates $r \geq 0.05$ we see again that convergence time decreases as both the evidence rate $r$ and network connectivity $k$ increases. We also see that for these higher evidence rates, increasing network connectivity $k$ to beyond $20$ nearest neighbours does not lead to further reductions in convergence time.
While \Cref{fig:convergence-time} demonstrates the ability of the model to reach a steady state under certain conditions, it does not demonstrate the learning accuracy of our model. To this end, we now present results showing the average error of the population according to \Cref{def:average-error}.

\begin{figure*}[t]
\begin{subfigure}{1\textwidth}
\centering
    \includegraphics[width=0.8\textwidth]{{er_legend_labelled}.pdf}
\end{subfigure}\hfill
\begin{subfigure}{0.33\textwidth}
    \includegraphics[width=1\textwidth]{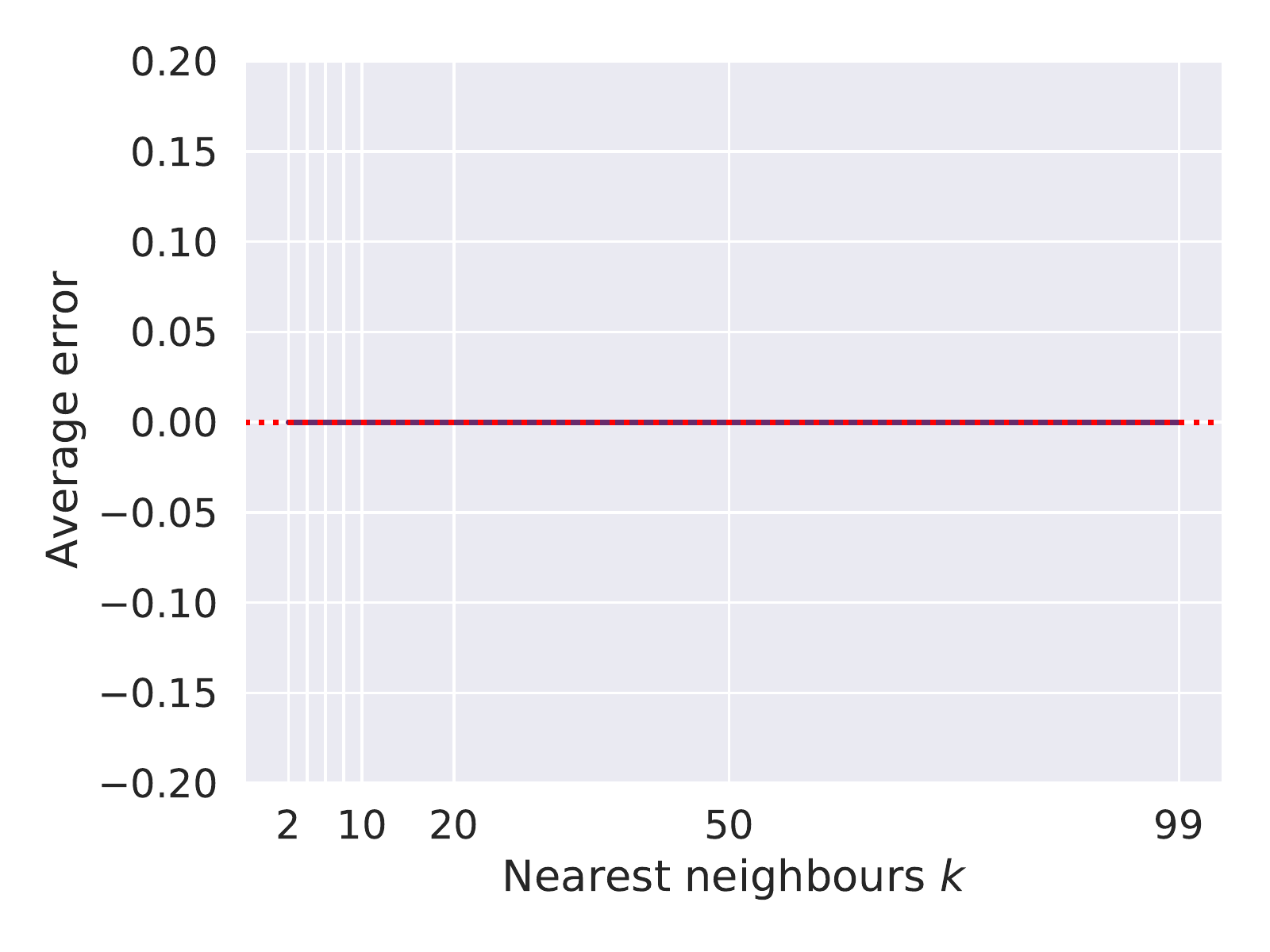}
    \subcaption{Noise $\epsilon = 0.0$.}
    \label{fig:er-noise-0}
\end{subfigure}\hfill
\begin{subfigure}{.33\textwidth}
    \includegraphics[width=1\textwidth]{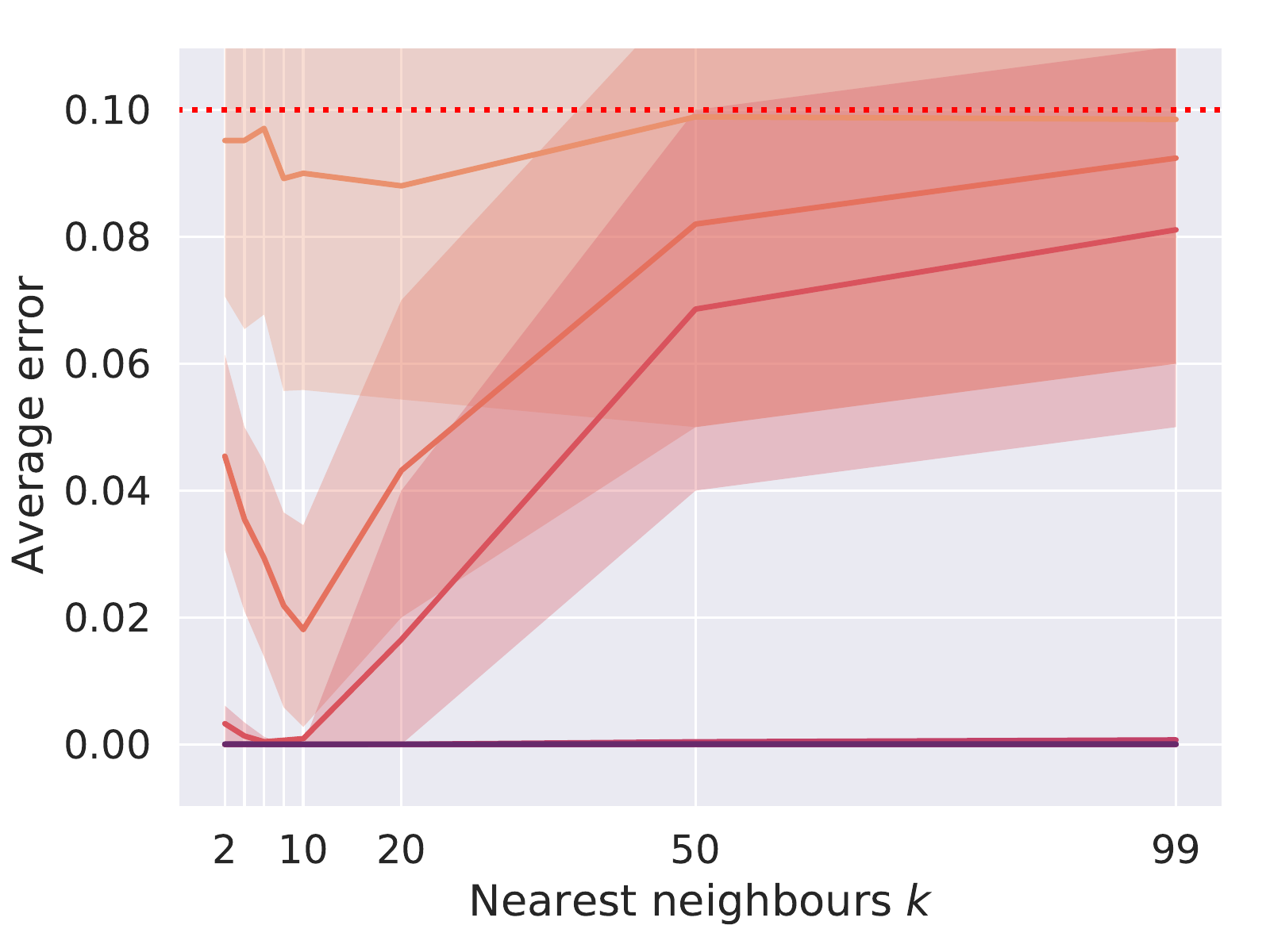}
    \subcaption{Noise $\epsilon = 0.1$.}
    \label{fig:er-noise-0.1}
\end{subfigure}\hfill
\begin{subfigure}{.33\textwidth}
    \includegraphics[width=1\textwidth]{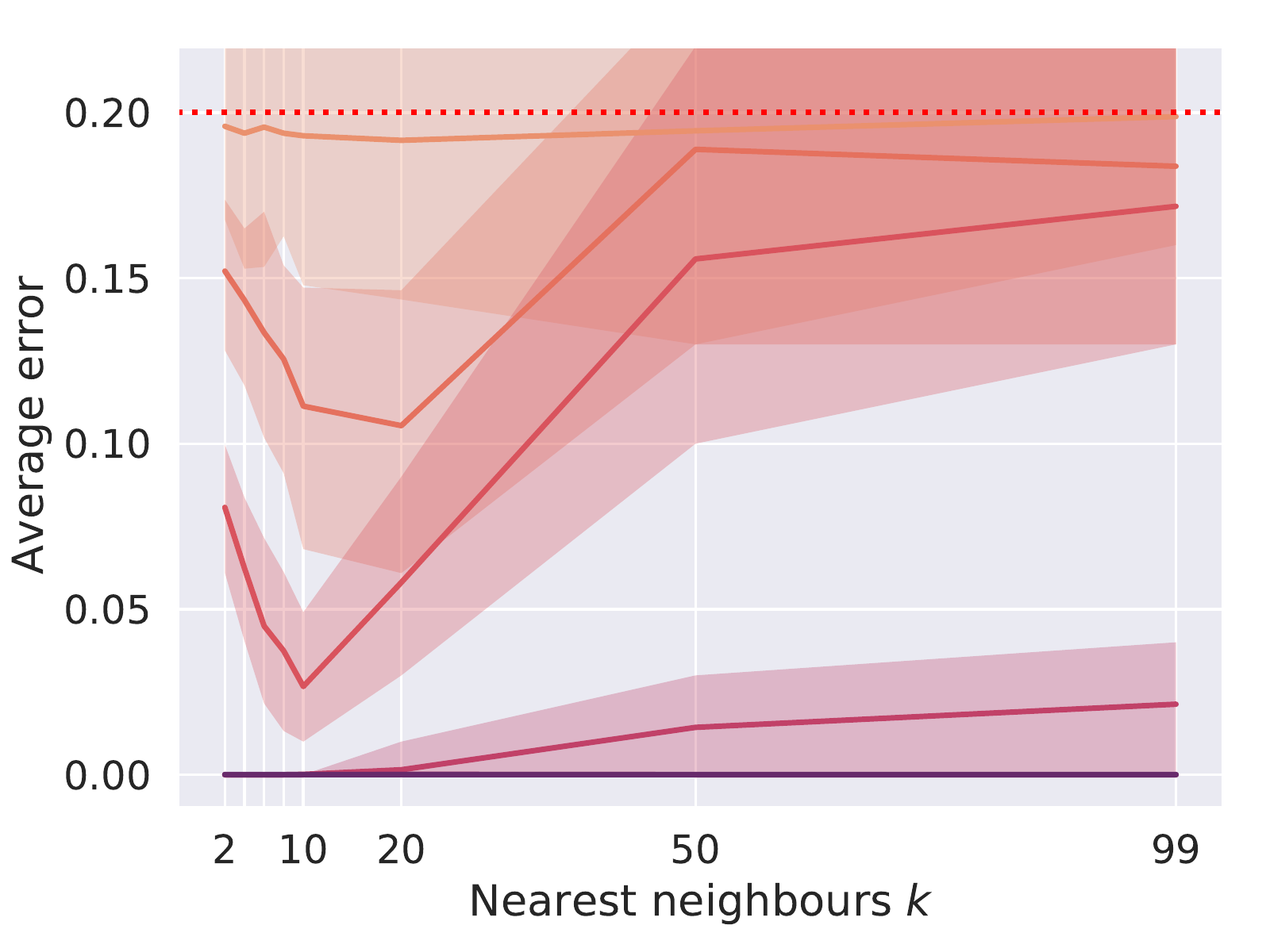}
    \subcaption{Noise $\epsilon = 0.2$.}
    \label{fig:er-noise-0.2}
\end{subfigure}\hfill
\begin{subfigure}{0.33\textwidth}
    \includegraphics[width=1\textwidth]{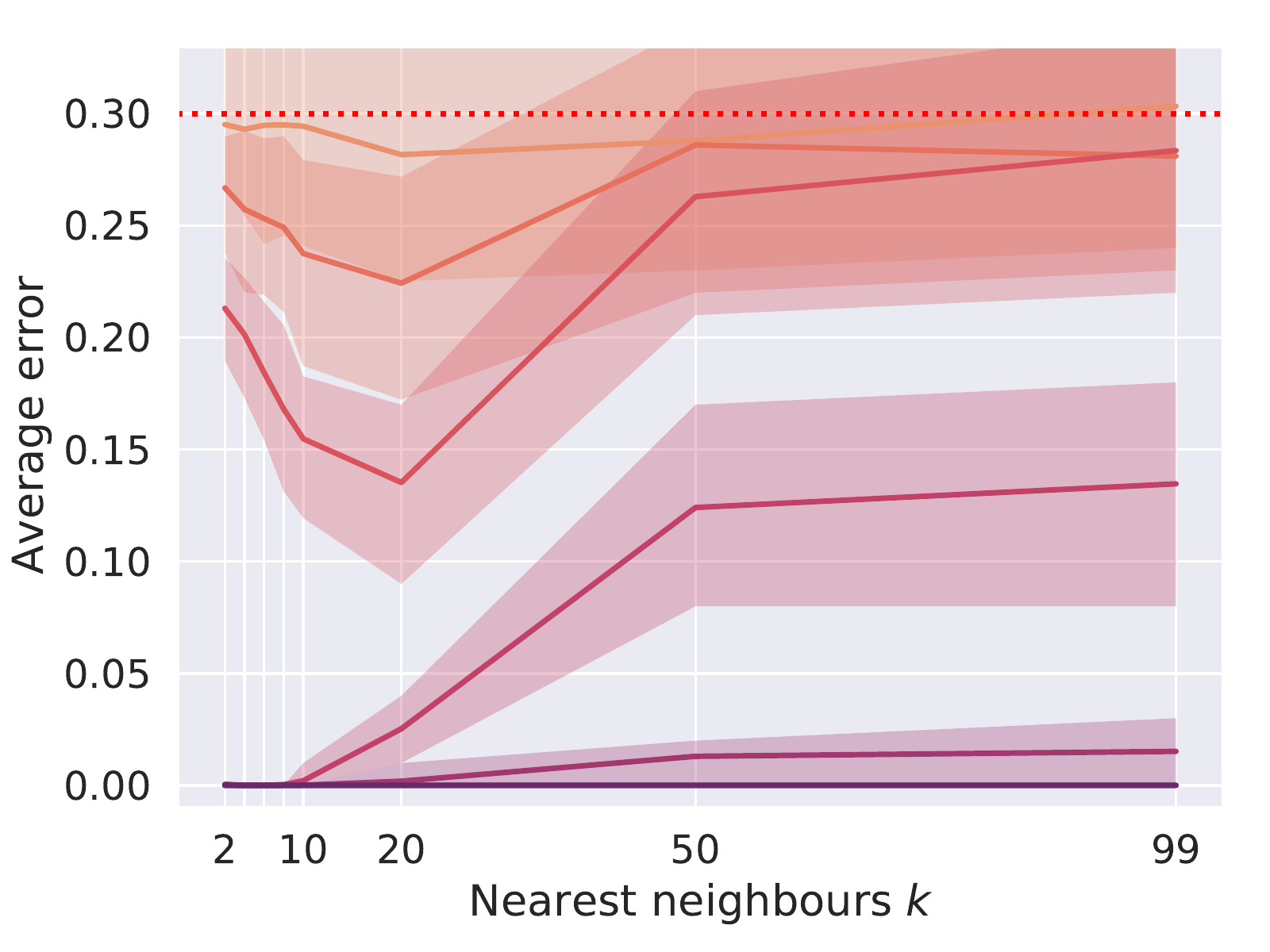}
    \subcaption{Noise $\epsilon = 0.3$.}
    \label{fig:er-noise-0.3}
\end{subfigure}\hfill
\begin{subfigure}{.33\textwidth}
    \includegraphics[width=1\textwidth]{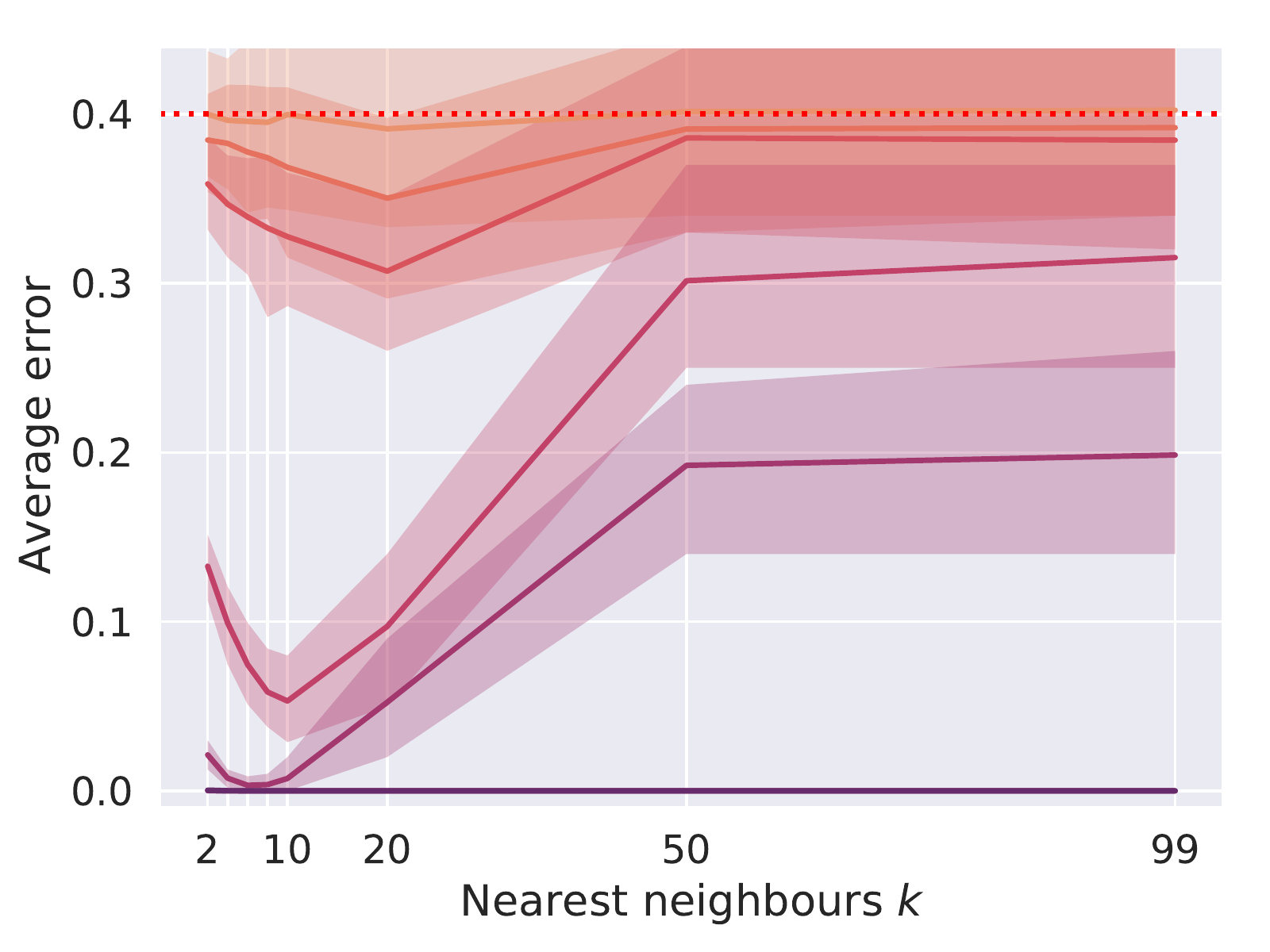}
    \subcaption{Noise $\epsilon = 0.4$.}
    \label{fig:er-noise-0.4}
\end{subfigure}\hfill
\begin{subfigure}{.33\textwidth}
    \includegraphics[width=1\textwidth]{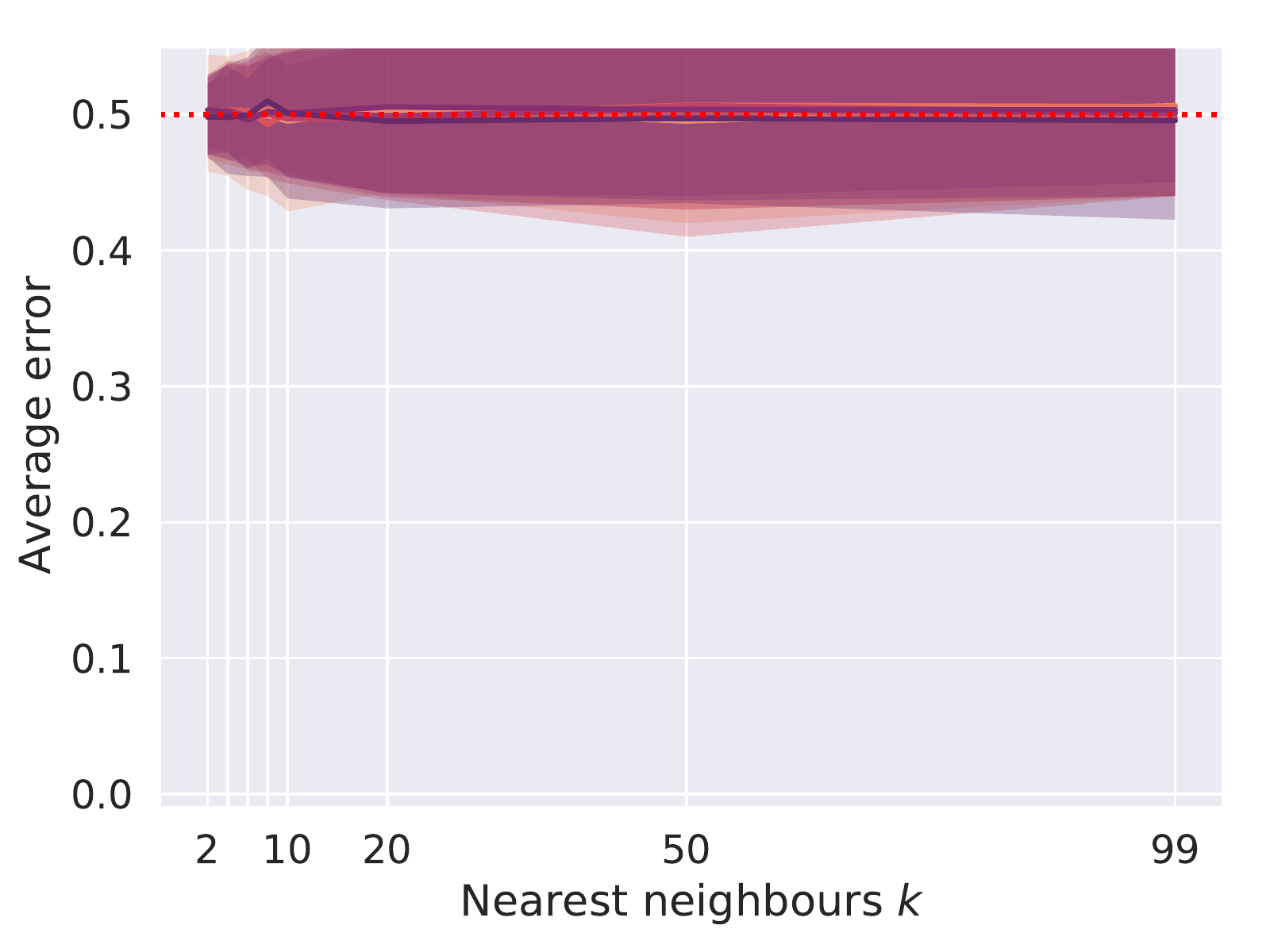}
    \subcaption{Noise $\epsilon = 0.5$.}
    \label{fig:er-noise-0.5}
\end{subfigure}\hfill
\caption{Average error of a population of $100$ agents for nearest neighbours $k$. Each solid line depicts a different evidence rate $r \in [0.01,1]$, while the red dotted line depicts the noise level $\epsilon$.}
\label{fig:error-steadystates}
\end{figure*}

In \Cref{fig:error-steadystates} we show the average error of the population at steady state against the number of nearest neighbours $k$ in the network. \Cref{fig:er-noise-0} shows that when the system is free from noise (i.e., when $\epsilon = 0$) the model always converges on the true state of the world with $0$ average error at steady state. This is true for all values of $k$ and all evidence rates $r$ including $r = 0.001$ which, for a population of $100$ agents, corresponds to the population receiving a single piece of evidence once every $10$ time steps on average. However, in a scenario in which evidence is always accurate, these results are to be expected.
Another edge case is when $\epsilon = 0.5$, resulting in agents always receiving random evidence. Unsurprisingly, as shown in \Cref{fig:er-noise-0.5}, the population always converges on an average error of around $\epsilon$ due to its inability to receive informative evidence for any particular proposition.

\Crefrange{fig:er-noise-0.1}{fig:er-noise-0.4} show the average error of the population for increasing levels of noise $\epsilon$ from $0.1$ to $0.4$. Compared with the noise-free scenario, the connectivity of the network has a clear impact on the learning accuracy of our model when agents encounter noisy evidence.
For all noise levels $\epsilon$ in this range we see that the model performs well with respect to accuracy as the population consistently achieves an average error at or below $\epsilon$ for all evidence rates $r$. While the population does not always learn the true state of the world, the average error of the population can be significantly reduced below $\epsilon$ by adopting a less connected network. Indeed, for a totally-connected network, i.e., where $k = 99$, the average error is very often greater than networks with lower connectivity.
In the extreme cases, with evidence rates $r \geq 0.5$ the population always learns the true state of the world for $\epsilon < 0.5$.

\begin{figure*}[t]
\begin{subfigure}{1\textwidth}
\centering
    \hspace{2em}\includegraphics[width=0.35\textwidth]{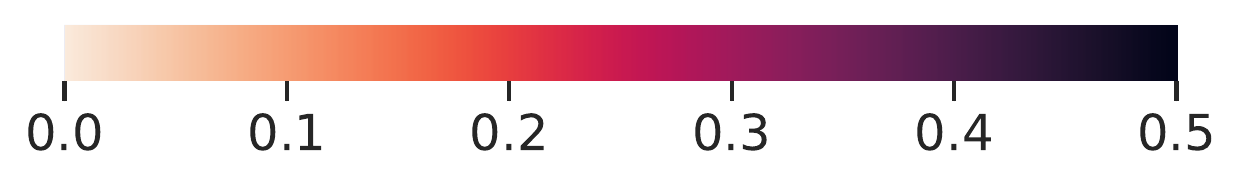}
\end{subfigure}\hfill
\begin{subfigure}{.33\textwidth}
    \includegraphics[width=1\textwidth]{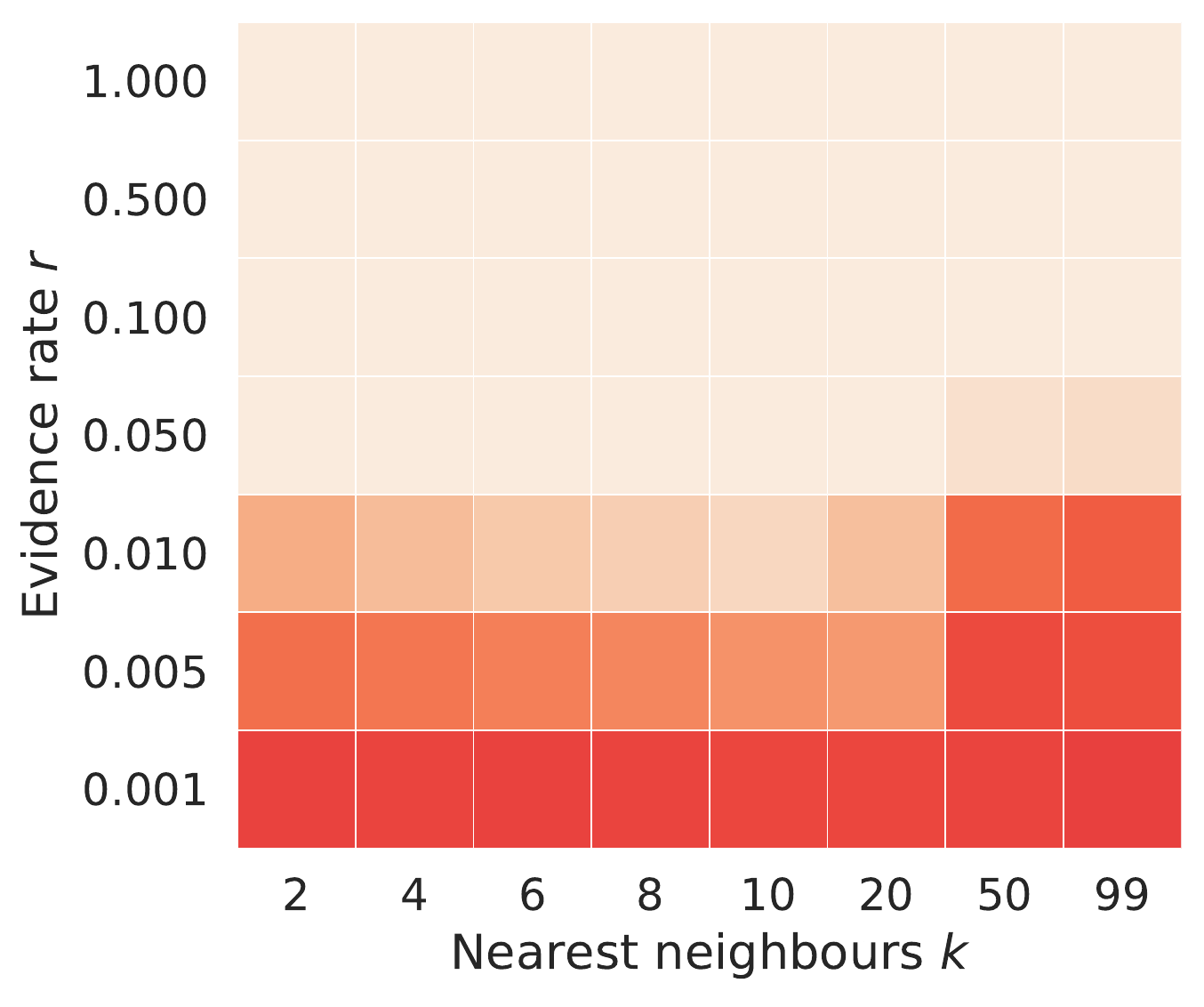}
    \subcaption{Noise $\epsilon = 0.2$.}
    \label{fig:hm-noise-0.2}
\end{subfigure}\hfill
\begin{subfigure}{0.33\textwidth}
    \includegraphics[width=1\textwidth]{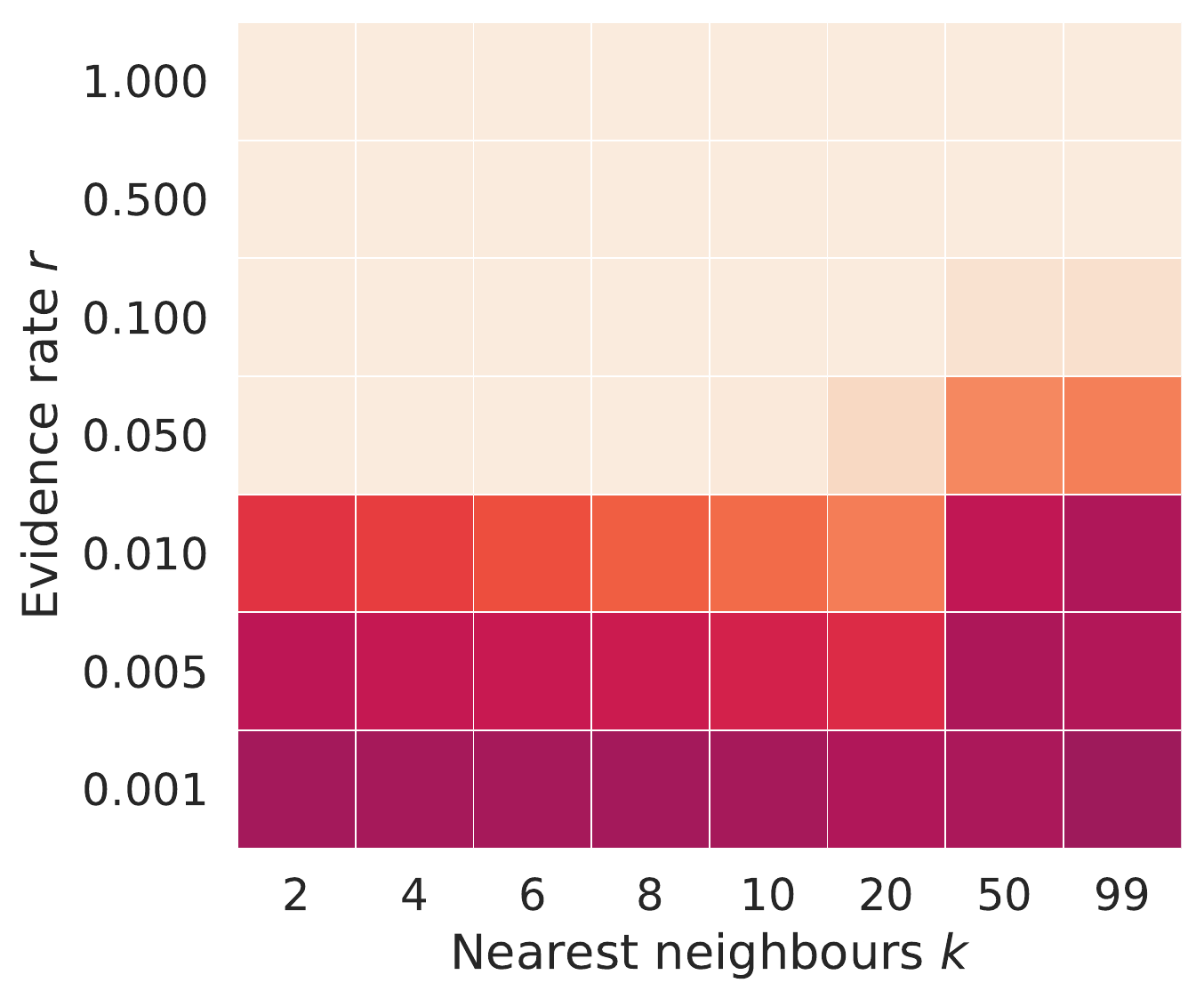}
    \subcaption{Noise $\epsilon = 0.3$.}
    \label{fig:hm-noise-0.3}
\end{subfigure}\hfill
\begin{subfigure}{.33\textwidth}
    \includegraphics[width=1\textwidth]{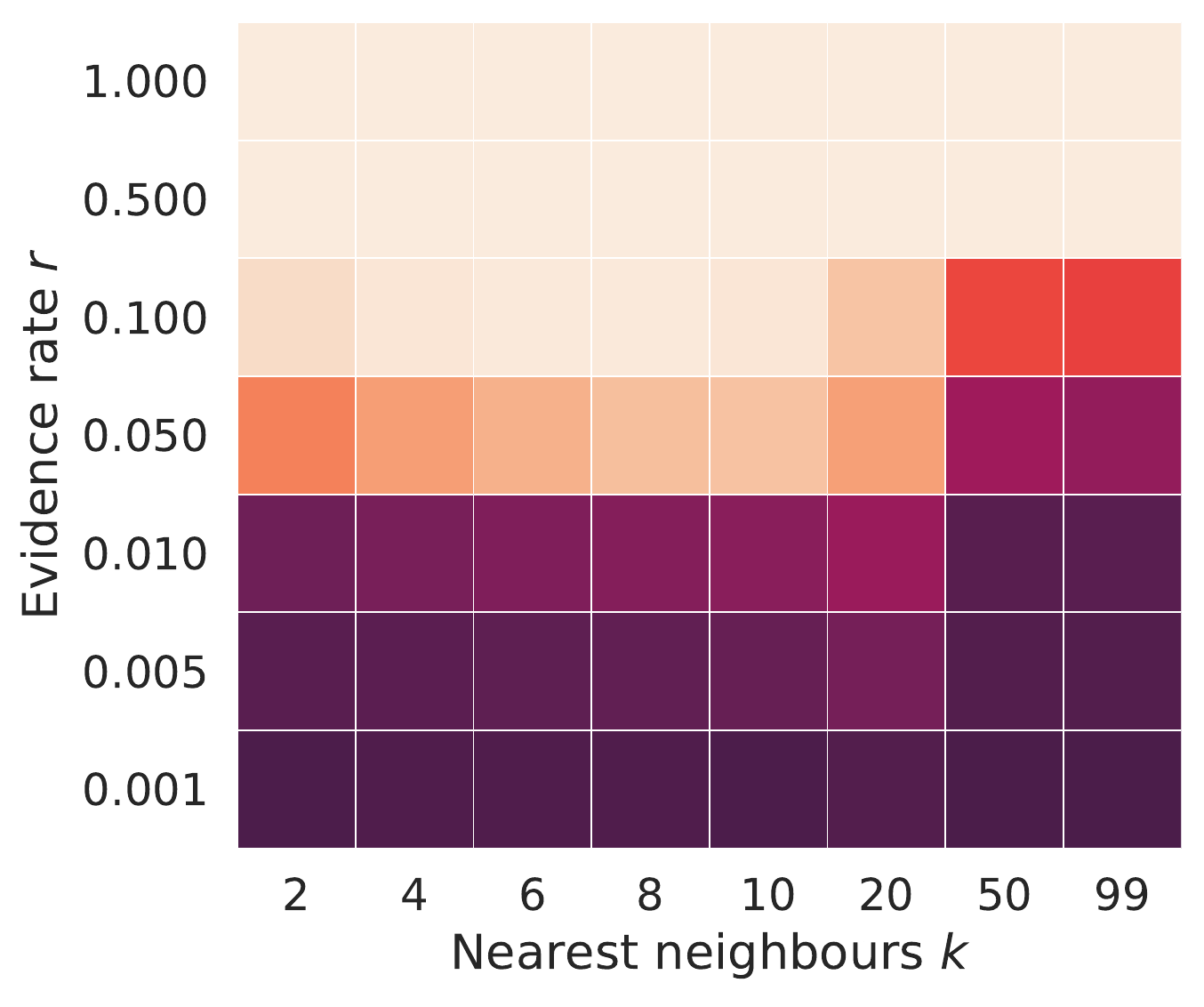}
    \subcaption{Noise $\epsilon = 0.4$.}
    \label{fig:hm-noise-0.4}
\end{subfigure}\hfill
\caption{Average error of a population of $100$ agents for both evidence rates $r \in [0.01,1]$ and nearest neighbours $k = 2, \dots, 99$.}
\label{fig:heatmaps}
\end{figure*}

Alternatively, in \Cref{fig:heatmaps} we show heatmaps of the average error at steady state for evidence rate $r$ and the number of nearest neighbours $k$. Focussing on a range of $\epsilon$ from $0.2$ to $0.4$ we see again that small-world networks with less connectivity outperform networks with greater connectivity, including totally-connected networks.
Broadly, given a noisy scenario and an environment with sparse evidence, there is a network with connectivity $k$ that outperforms networks of both lesser and greater connectivity. For environments with higher evidence rates $r$, a less connected small-world network improves the accuracy of collective learning.

\subsection{Convergence results for small-world networks with random rewiring}
\label{ssec:random-connectivity}

\begin{figure*}[t]
\begin{subfigure}{1\textwidth}
\centering
    \hspace{2em}\includegraphics[width=0.8\textwidth]{{er_legend_labelled}.pdf}
\end{subfigure}\hfill
\begin{subfigure}{.33\textwidth}
    \includegraphics[width=1\textwidth]{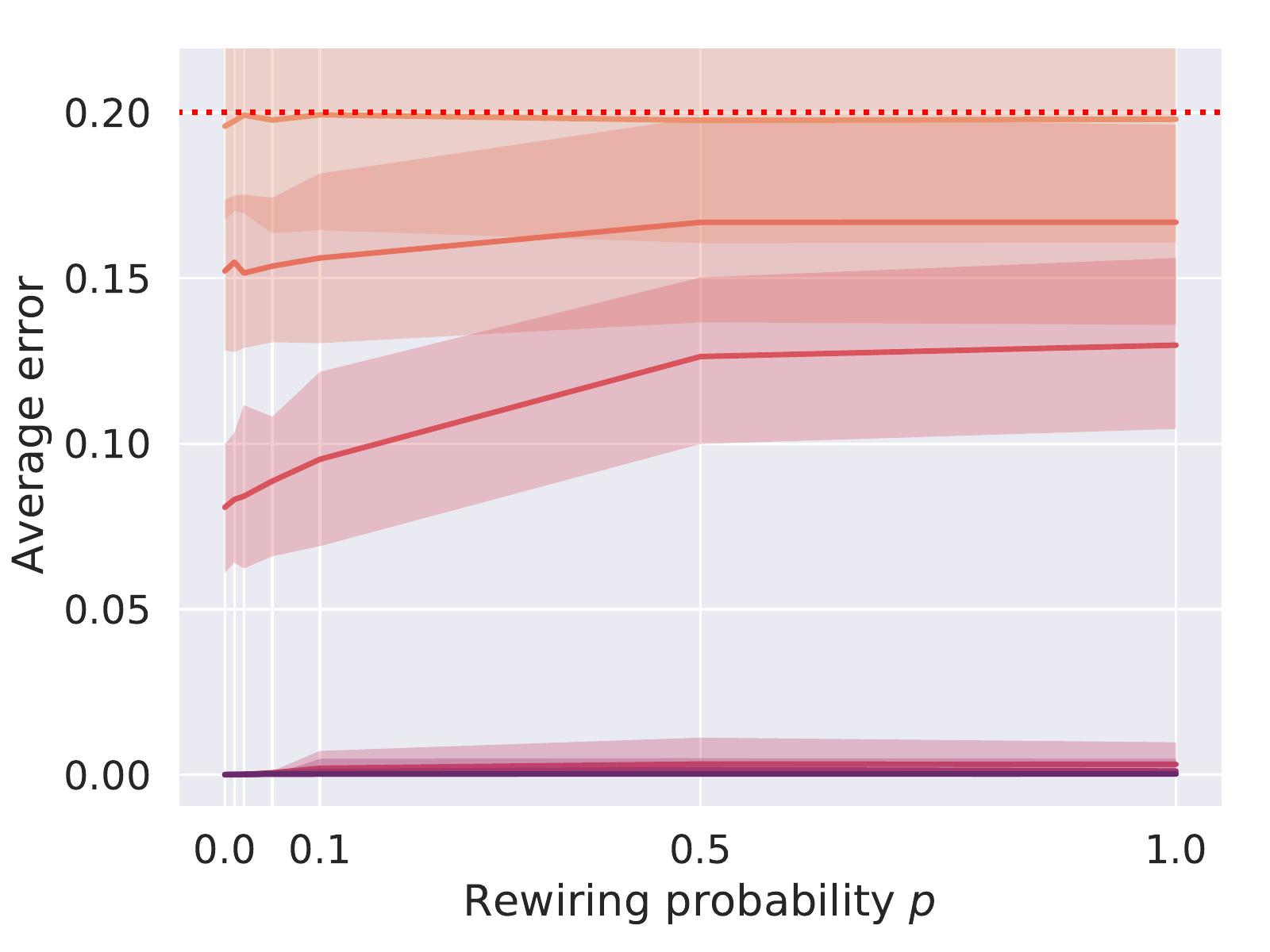}
    \subcaption{$k = 2$.}
    \label{fig:rewiring-k-2}
\end{subfigure}\hfill
\begin{subfigure}{.33\textwidth}
    \includegraphics[width=1\textwidth]{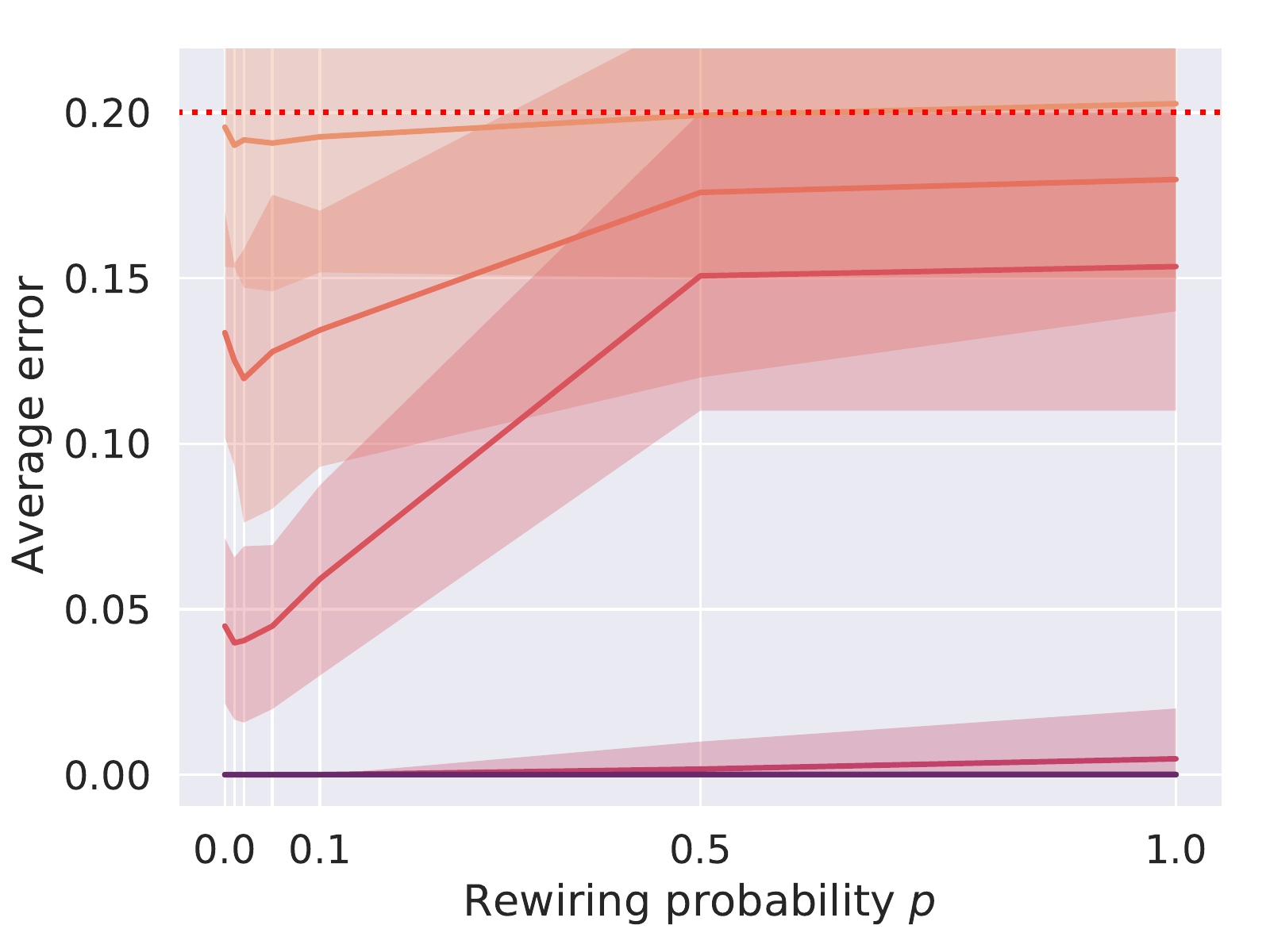}
    \subcaption{$k = 6$.}
    \label{fig:rewiring-k-6}
\end{subfigure}\hfill
\begin{subfigure}{.33\textwidth}
    \includegraphics[width=1\textwidth]{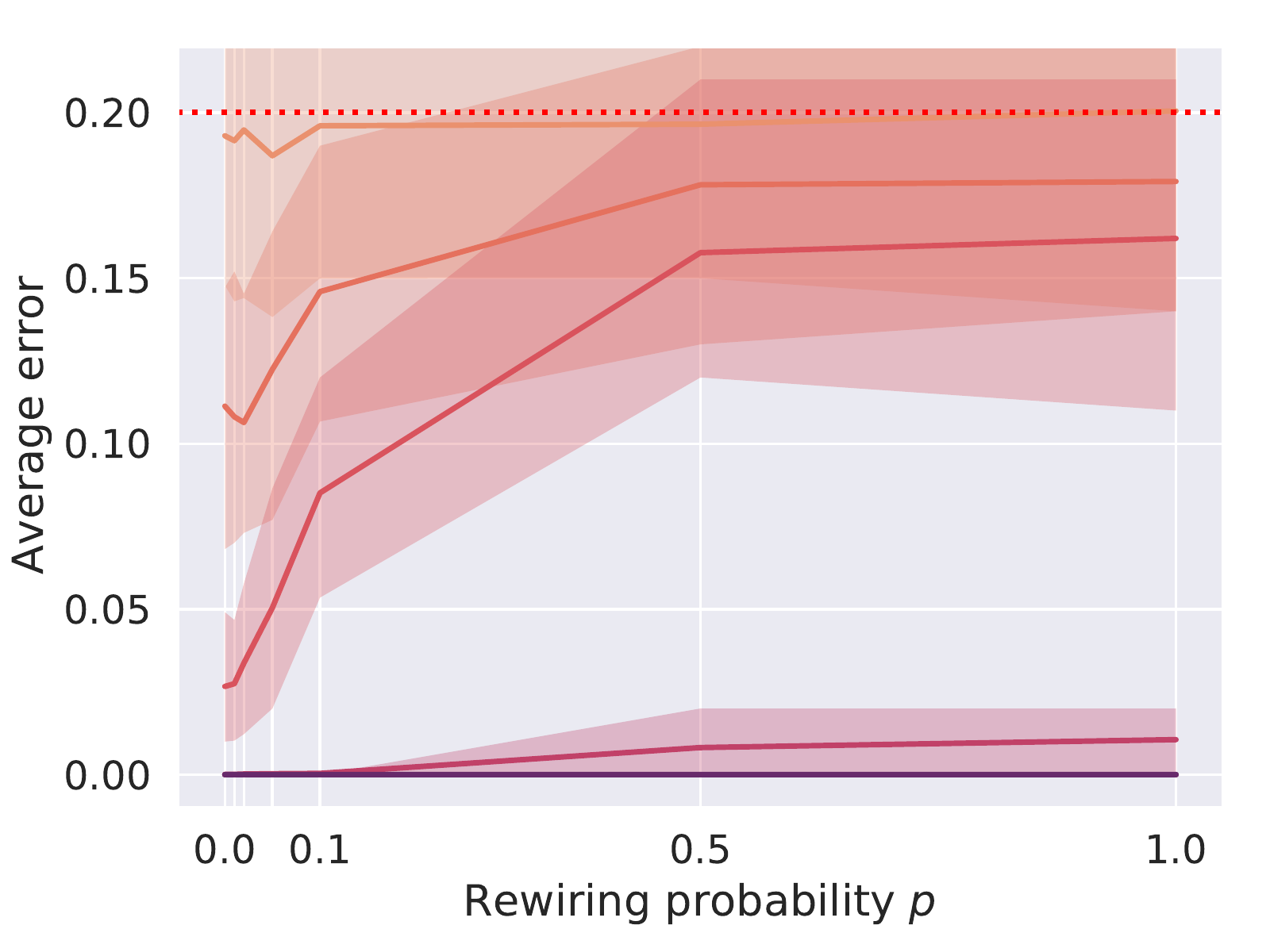}
    \subcaption{$k = 10$.}
    \label{fig:rewiring-k-10}
\end{subfigure}\hfill
\caption{Average error of a population of $100$ agents against rewiring probability $\rho \in [0,1]$. Each solid line depicts a different evidence rate $r \in [0.01,1]$ with noise $\epsilon = 0.2$, while the red dotted line depicts the noise level $\epsilon$.}
\label{fig:rewiring-steadystates}
\end{figure*}

Having studied the primary parameter $k$ of small-world networks in \Cref{ssec:results}, we now study the secondary parameter associated with small-world networks: the rewiring probability $\rho$. This randomising parameter reduces the regularity of small-world networks by rewiring connections between neighbouring nodes to agents of greater separation in the network. The purpose of this parameter is to introduce additional connections (or `paths') in the network that, for small values of $k$, are likely to improve information propagation in the network.

\Cref{fig:rewiring-steadystates} shows the average error of the population at steady state with moderate noise $\epsilon = 0.2$ and for different rewiring probabilities $\rho$ between $0$ and $1$. In \Cref{fig:rewiring-k-2} we see that for $k = 2$, corresponding to a ring network in which agents are connected to their two nearest neighbours only, the average error of the population increases with $\rho$ for $r$ from $0.005$ to $0.05$ while remaining stable for very low and very high evidence rates, i.e., for $r = 0.001$ and $r \geq 0.1$.
For example, with $r = 0.01$, when the network is totally regular with $\rho = 0$, the average error is $0.081$, while for $\rho = 1$, a network with totally random connectivity, the average error increases by $60\%$ to $0.130$.
The same effect is broadly observed in \Cref{fig:rewiring-k-6,fig:rewiring-k-10} for the same range of evidence rates, except that random rewiring increases the average error of the population to a greater extent in small-world networks with greater connectivity. For example, in \Cref{fig:rewiring-k-10} with $k = 10$ and an evidence rate $r = 0.01$, when the network is totally regular (i.e., when $\rho = 0$) the average error of the population is $0.027$. When $\rho = 0.1$, the average error increases to $0.085$ while for total randomness (i.e., when $\rho = 1$) the average error is $0.162$ which is a $600\%$ increase in average error compared with the $\rho = 0$ case.

As the network connectivity is altered from total regularity to total randomness, i.e., from $\rho = 0$ to $1$, respectively, the population consistently performs worse when attempting to learn the true state of the world, with the average error of the population increasing as $\rho$ increases. It is clear, therefore, that irregularity in the network negatively impacts performance in a collective learning setting.

\section{Discussion and Conclusion}
\label{sec:conclusion}

In this paper we have investigated the importance of considering the connectivity of the underlying network for collective learning in autonomous systems. The environment is represented by a set of descriptors in the form of propositional variables and agents' beliefs are represented by an allocation of three-valued truth values to each of the propositions. Agents adopt a combined process of evidential updating, learning directly from the environment, and belief fusion, combining their beliefs with other agents to form a pairwise consensus while correcting for inconsistencies that have arisen from noisy evidence. In this context, we have studied how the structure of the underlying network impacts the dynamics of a system of $100$ agents learning the state of the world for a range of scenarios with varying levels of evidence sparsity and noise.

We have shown that a less-connected small-world network leads to greater accuracy on a collective learning task when compared with a totally-connected network. Through simulation studies our results show that, when the evidence in the environment is both sparse and noisy, then a network with moderate connectivity $k$ outperforms networks with lower or higher connectivity for some combination of evidence rate $r$ and noise level $\epsilon$.
Broadly, the optimal level of connectivity is always lower than total connectivity when the underlying network retains high regularity, i.e., $\rho \approx 0$. As the network connectivity becomes increasingly random, i.e., as $\rho \rightarrow 1$, the accuracy of the system decreases.

% In a collective learning setting when the process of gathering evidence is subject to noise, a totally regular, moderately connected small-world network, e.g.,for $r = 0.01$ a network with $k = 10$ and $\rho = 0$, results in the best performance.

\section*{Acknowledgements}
\label{sec:acknowledgements}

This work was funded and delivered in partnership between Thales Group, University of Bristol and with the support of the UK Engineering and Physical Sciences Research Council, ref. EP/R004757/1 entitled ``Thales-Bristol Partnership in Hybrid Autonomous Systems Engineering (T-B PHASE).''

\bibliographystyle{spmpsci}
\bibliography{references}

\begin{thebibliography}{10}
\providecommand{\url}[1]{{#1}}
\providecommand{\urlprefix}{URL }
\expandafter\ifx\csname urlstyle\endcsname\relax
  \providecommand{\doi}[1]{DOI~\discretionary{}{}{}#1}\else
  \providecommand{\doi}{DOI~\discretionary{}{}{}\begingroup
  \urlstyle{rm}\Url}\fi

\bibitem{Balenzuela}
Balenzuela, P., Pinasco, J.P., Semeshenko, V.: The undecided have the key:
  Interaction-driven opinion dynamics in a three state model.
\newblock PLOS ONE \textbf{10}(10), 1--21 (2015)

\bibitem{Baronchelli}
Baronchelli, A.: The emergence of consenus: a primer.
\newblock Royal Society Open Science \textbf{5}(2), 172,189 (2018)

\bibitem{Brambilla}
Brambilla, M., Ferrante, E., Birattari, M., Dorigo, M.: Swarm robotics: a
  review from the swarm engineering perspective.
\newblock Swarm Intelligence \textbf{7}(1), 1--41 (2013)

\bibitem{Crosscombe2016}
Crosscombe, M., Lawry, J.: A model of multi-agent consensus for vague and
  uncertain beliefs.
\newblock Adaptive Behavior \textbf{24}(4), 249--260 (2016)

\bibitem{Crosscombe2017}
{Crosscombe}, M., {Lawry}, J., {Hauert}, S., {Homer}, M.: Robust distributed
  decision-making in robot swarms: Exploiting a third truth state.
\newblock In: 2017 IEEE/RSJ International Conference on Intelligent Robots and
  Systems (IROS), pp. 4326--4332 (2017)

\bibitem{Dabarera2019}
{Dabarera}, R., {Wickramarathne}, T.L., {Premaratne}, K., {Murthi}, M.N.:
  Achieving consensus under bounded confidence in multi-agent distributed
  decision-making.
\newblock In: 2019 22th International Conference on Information Fusion
  (FUSION), pp. 1--8 (2019)

\bibitem{DeGroot}
DeGroot, M.H.: Reaching a consensus.
\newblock Journal of the American Statistical Association \textbf{69}(345),
  118--121 (1974)

\bibitem{Douven2019}
Douven, I.: Optimizing group learning: An evolutionary computing approach.
\newblock Artificial Intelligence \textbf{275}, 235 -- 251 (2019)

\bibitem{Douven2011}
Douven, I., Kelp, C.: Truth approximation, social epistemology, and opinion
  dynamics.
\newblock Erkenntnis \textbf{75}(2), 271 (2011)

\bibitem{Erdos1959}
Erd\"{o}s, P., R\'{e}nyi, A.: On random graphs.
\newblock Publicationes Mathematicae (Debrecen) \textbf{6}, 290--297 (1959)

\bibitem{Franks}
Franks, N., Pratt, S., Mallon, E., Britton, N., Sumpter, D.: Information flow,
  opinion-polling and collective intelligence in house-hunting social insects.
\newblock Philosophical Transactions B: Biological Sciences \textbf{357
  (1427)}, 1567 -- 1583 (2002)

\bibitem{Hegselmann2005}
Hegselmann, R., Krause, U.: Opinion dynamics driven by various ways of
  averaging.
\newblock Computational Economics \textbf{25}(4), 381--405 (2005)

\bibitem{Hegselmann2002}
Hegselmann, R., Krause, U., et~al.: Opinion dynamics and bounded confidence
  models, analysis, and simulation.
\newblock Journal of artificial societies and social simulation \textbf{5}(3)
  (2002)

\bibitem{Lawry2019}
Lawry, J., Crosscombe, M., Harvey, D.: Epistemic sets applied to best-of-n
  problems.
\newblock In: G.~Kern-Isberner, Z.~Ognjanovi{\'{c}} (eds.) Symbolic and
  Quantitative Approaches to Reasoning with Uncertainty, pp. 301--312. Springer
  International Publishing, Cham (2019)

\bibitem{Lee2018}
Lee, C., Lawry, J., Winfield, A.: Negative updating combined with opinion
  pooling in the best-of-n problem in swarm robotics.
\newblock In: Swarm Intelligence, pp. 97--108. Springer International
  Publishing, Cham (2018)

\bibitem{Masuda}
Masuda, N., Aihara, K.: Global and local synchrony of coupled neurons in
  small-world networks.
\newblock Biological cybernetics \textbf{90}, 302--9 (2004)

\bibitem{Newman}
Newman, M.E.J., Watts, D.J., Strogatz, S.H.: Random graph models of social
  networks.
\newblock Proceedings of the National Academy of Sciences \textbf{99}(suppl 1),
  2566--2572 (2002)

\bibitem{Olfati-Saber}
{Olfati-Saber}, R., {Fax}, J.A., {Murray}, R.M.: Consensus and cooperation in
  networked multi-agent systems.
\newblock Proceedings of the IEEE \textbf{95}(1), 215--233 (2007)

\bibitem{Parker}
Parker, C.A.C., Zhang, H.: {Cooperative Decision-Making in Decentralized
  Multiple-Robot Systems: The Best-of-N Problem}.
\newblock IEEE/ASME Transactions on Mechatronics \textbf{14}(2), 240--251
  (2009)

\bibitem{Parker2011}
Parker, C.A.C., Zhang, H.: Biologically inspired collective comparisons by
  robotic swarms.
\newblock The International Journal of Robotics Research \textbf{30}(5),
  524--535 (2011)

\bibitem{Perron}
Perron, E., Vasudevan, D., Vojnovi{\'{c}}, M.: {Using three states for binary
  consensus on complete graphs}.
\newblock Proceedings - IEEE INFOCOM pp. 2527--2535 (2009)

\bibitem{Rubenstein}
{Rubenstein}, M., {Ahler}, C., {Nagpal}, R.: Kilobot: A low cost scalable robot
  system for collective behaviors.
\newblock In: 2012 IEEE International Conference on Robotics and Automation,
  pp. 3293--3298 (2012)

\bibitem{Schranz}
Schranz, M., Umlauft, M., Sende, M., Elmenreich, W.: Swarm robotic behaviors
  and current applications.
\newblock Frontiers in Robotics and AI \textbf{7}, 36 (2020)

\bibitem{Stone}
Stone, M.: The opinion pool.
\newblock The Annals of Mathematical Statistics \textbf{32}(4), 1339--1342
  (1961)

\bibitem{Valentini2017}
Valentini, G., Ferrante, E., Dorigo, M.: The best-of-n problem in robot swarms:
  Formalization, state of the art, and novel perspectives.
\newblock Frontiers in Robotics and AI \textbf{4}, 9 (2017)

\bibitem{Valentini2015}
Valentini, G., Hamann, H., Dorigo, M.: Efficient decision-making in a
  self-organizing robot swarm: On the speed versus accuracy trade-off.
\newblock In: Proceedings of the 2015 International Conference on Autonomous
  Agents and Multiagent Systems, AAMAS '15, p. 1305–1314. International
  Foundation for Autonomous Agents and Multiagent Systems, Richland, SC (2015)

\bibitem{Watts2003}
Watts, D.J.: Small Worlds: The Dynamics of Networks between Order and
  Randomness.
\newblock Princeton University Press, USA (2003)

\bibitem{Watts1998}
Watts, D.J., Strogatz, S.H.: {Collective dynamics of ‘small-world' networks}.
\newblock Nature \textbf{393}(June), 440--442 (1998)

\end{thebibliography}

\end{document}